\documentclass[showpacs,10pt,aps,prd,reprint,onecolumn,notitlepage,nofootinbib,superscriptaddress]{revtex4-1}

\usepackage{hyperref}
\usepackage{amsmath}
\usepackage{mathtools}
\usepackage{bm}
\usepackage{amssymb}
\usepackage{graphicx}
\usepackage[]{subfigure}
\usepackage[usenames,dvipsnames,svgnames,table]{xcolor}
\usepackage{filecontents}

\hypersetup{
    bookmarks=true,         
    pdftoolbar=true,        
    pdfmenubar=true,        
    pdffitwindow=true,     
    pdfstartview={FitH},     
    pdftitle={My title},     
    pdfauthor={author},      
    pdfsubject={Subject},    
    pdfcreator={Creator},    
    pdfproducer={Producer},  
    pdfkeywords={keyword1} 
    pdfnewwindow=true,       
    colorlinks=true ,        
    linkcolor=blue  ,        
    citecolor=blue,      
    filecolor=green,       
    urlcolor=blue            
}

\providecommand{\ed}{\mathrm{d}}
\providecommand{\bu}{{\bm{u}}}
\providecommand{\bp}{{\bm{p}}}
\providecommand{\bJ}{{\bm{J}}}
\providecommand{\ba}{{\bm{a}}}

\begin{document}

\title{Homothetic Congruences in General Relativity}

\author{Mohsen Fathi}
\email{m.fathi@shargh.tpnu.ac.ir; \,\,mohsen.fathi@gmail.com}

 \affiliation{Department of Physics, 
Payame Noor University (PNU),
P.O. Box 19395-3697 Tehran, Iran}

\begin{abstract}
The kinematical characteristics of distinct infalling homothetic fields are discussed by specifying the transverse subspace of their generated congruences to the energy-momentum deposit of the chosen gravitational system. This is pursued through the inclusion of the base manifold's cotangent bundle in a generalized Raychaudhuri equation and its kinematical expressions. Exploiting an electromagnetic energy-momentum tensor as the source of non-gravitational effects, I investigate the evolution of the mentioned homothetic congruences, as they fall onto a Reissner-Nordstr\"{o}m black hole. The results show remarkable differences to the common expectations from infalling congruences of massive particles.  

\bigskip

\noindent{\textit{keywords}}:  Homothetic fields, Generalized Raychaudhuri equation, Congruence kinematics, Cotangent bundle
\end{abstract}

\pacs{04.20.-q, 04.20.Dw, 04.20.Jb, 04.50.Kd} 
\maketitle

\section{Introduction}

The importance of the particles' geodesics, revealed by the advent of general relativity. In fact, this theory on its own, stems in the consideration of geodesics of freely falling objects in a gravitational field, which constitutes a crucial link with Riemannian geometry \cite{MTW1973}. These geodesics were used practically to make mathematical predictions of astronomical phenomena, as the foundations of the observational tests of general relativity. The famous observation of Mercury's perihelon shift, by Eddington, is a well-known example. While the geodesics are usually regarded as the path of single particles falling freely in gravitational fields, they can become of more interest when considered as a set; a {\textit{congruence}}\footnote{It is said that the word {\textit{congruence}} was first used by Gauss \cite{Gauss1986}.}. Geodesic congruences are of great importance in general relativity, because they give us information about the evolution of the kinematics of falling objects in the geometry defined by black holes as well as telling about the standard cosmological dust \cite{Hawking1973}. They also constitute an important tool to inspect the Penrose-Hawking singularity theorems \cite{Penrose1965,Hawking1965,Hawking1966,Penrose2002}. This tool has indeed been formulated by the well-known Raychauduri equation which was firstly aimed at studying the evolution of the cosmic fluid \citep{Raychaudhuri1955}. Essentially, the Raychaudhuri equation is a geometrical representation, to govern the kinematical decomposition of congruences and reveals their evolution. In fact, this equation tells us about how the congruences would diverge, converge or twist, by calculating the changes of their cross-sectional (transverse) area \cite{Poisson2009}(for reviews, see also Refs. \cite{Ellis2007,Kar2007}). The Raychaudhuri equation can also be applied to discuss the energy conditions and therefore, on the attractiveness/repulsiveness of gravity which is entangled with the concept of geodesic focusing. This equation is also used to study the solutions of general relativity \cite{Stephani2003}. It is therefore apparent that why the geodesic congruences have been of great interest for several decades after the birth of general relativity.

That said, I should note that since not all the congruences are geodesic, the congruence kinematics may be affected by the presence of an acceleration. This can be regarded as both geometrical and physical. Whereas the former may depend on the specific choice of observers and is therefore completely observer-dependent, the latter is a consequence of the presence of non-gravitational sources (like an electrically charged mass). The non-geodesic property would therefore provide a sensible subject to explore, since it can have assessable correspondents in the nature. Such property may vary from a congruence which is not affinely parameterized, to that which is evolving in an electromagnetic field.

\subsection{The Aim of This Paper}

In this study, I will consider both the above cases. In fact, I take the congruences to have a natural acceleration. As mentioned above, the importance of investigating accelerated falling objects in gravitational fields, is in their valid analogs in the nature. To elaborate this in the current study, I investigate the properties of {\textit{homothetic}} non-null congruences in the presence of the electromagnetic energy-momentum tensor in the context of general relativity. In fact, homothetic vector fields have received remarkable attention form the community. Beside their application in finding exact solutions to Einstein field equations \cite{Stephani2003}, the homothetic fields are also of benefit in studying singularities in general relativity \cite{Hall1988}. In this regard, they have a considerable pertinence to my current study, because gravitational singularities are those regions where congruence convergence could happen\footnote{In this study, non-null congruences are examined. However, the singularities and their relevant cosmic censorship \cite{Penrose1999,Penrose2002} are usually discussed by means of null congruence expansion which enables us to discuss the evolution of black hole apparent horizons (see Ref.~\cite{Faraoni2015} for reviews).}. In a group theoretical approach, the spacetime symmetries can also be discussed by means of homothetic vector fields \cite{Hall2004,Ahmad2017}. To address further applications, its is worth noting their ability to classify cosmological models in general relativity and alternative gravity \cite{Shabbir2010,Shabbir2015,Ali2016,Shabbir2017}. 

In this paper however, I am interested in scrutinizing the kinematical characteristics of homothetic flows as they evolve in gravitational fields. In fact, homothetic motions in the geometrical sense, have been of great interest because they can talk about trivial and non-trivial space symmetries. Geometrical homothetic motions were discussed in details in Ref.~\citep{Yano1955} with subsequent published papers discussing them in the general relativistic context (see for example Refs. \cite{McIntosh1976,Berger1976} and references therein). Recent research also include cosmological applications in both general relativity and alternative gravity (see for example Refs. \cite{Jaen2014,Gad2015}). However, as mentioned above, my aim is to discuss the motion of homothetic fields, when they appear as congruences. To enhance the discussion, let me introduce a homothetic vector field $\bm u$,  in a spacetime manifold $(\mathcal{M}, g_{\alpha\beta})$, as the congruence generator\footnote{Throughout this paper, bold fonts are used to denote 4-(co)vectors. Furthermore, $";"$ and $","$ stand respectively for covariant and partial differentiations. Moreover, overdots indicate $\frac{D}{\ed\tau}$ where $\tau$ is taken to be the curve parameterization on $\mathcal{M}$. I adopt a $(- + + +)$ sign convention and all indices are 4-dimensional.}. This vector field, obeys the condition \cite{Poisson2009}
\begin{equation}\label{eq:I}
 \mathcal L_{\bu} g_{\alpha\beta} = 2\xi g_{\alpha\beta}
\end{equation}
in which $\mathcal{L}_{\bm{X}}$ indicate the Lie differentiation along a vector field $\bm{X}$ and $\xi$ is a real constant. Equation (\ref{eq:I}) can be recast as
\begin{equation}\label{eq:II}
u_{\alpha ;\beta }+u_{\beta ;\alpha }=2 \xi  g_{\alpha \beta },
\end{equation}
from which, the acceleration (co)vector $\dot\bu\equiv\ba=\nabla_{\bu}\bu$ can be obtained as
\begin{equation}\label{eq:III}
a^{\alpha }=-\frac{1}{2}g^{\alpha \beta } (\bu\cdot\bu)_{;\beta }+2 \xi  u^{\alpha },
\end{equation}
where I have notated $g_{\alpha \beta } u^{\alpha } u^{\beta }\equiv \bu\cdot\bu$. The above relation, talks about the extent to which, the congruence generator can deviate from being parallel transported. Note that, for $\bu\cdot\bu=\mathrm{const.}$ (normalized congruence generators), we get back to the famous shape $\ba = 2 \xi  \bu$ which is the characteristic for non-affinely-parameterized geodesic congruences. Homothetic congruences also prove to have non-orthogonal deviation fields. For any congruence deviation vector field $\bJ$, also known as the Jacobi field, obeying the condition \cite{Poisson2009}
\begin{equation}\label{eq:IV}
\mathcal{L}_\bu\bm J = \mathcal{L}_\bJ\bu,
\end{equation}
it is proved that
\begin{equation}\label{eq:V}
(\dot{\bJ\cdot\bu})=2 \xi(\bJ\cdot\bu)
\end{equation}
holds for homothetic fields (see appendix \ref{app:appendixA}). This shows that whether or not  $\bu\cdot\bu=\mathrm{const.}$, $\bJ\cdot\bu\neq \mathrm{const.}$ holds in general and unlike geodesic congruences, the deviation fields are not orthogonal to the congruence generators. It can also be shown that the congruence deviation relation (Jacobi equation) for homothetic congruences reads as (see appendix \ref{app:appendixA})
\begin{equation}\label{eq:VI}
\ddot{J}^{\alpha }=-\frac{1}{2}g^{\alpha \lambda } (\bu\cdot\bu)_{,\lambda \beta } J^\beta + 2 \xi  \dot{J}^{\alpha }-R^{\alpha }{}_{\mu \beta \gamma } u^{\mu } J^{\beta } u^{\gamma },
\end{equation}
indicating the contribution of the rate of change of the deviation field along the congruence, even for normalized generators. 

In this paper, such congruence generators are subjected to the effects of an electrically charged massive source of gravity. This way, they could show interesting behaviors because at the same time, the congruence is endowed with a geometric acceleration and a physical acceleration due to a non-gravitational energy. These would certainly affect the congruence evolution. Since the congruence kinematics are to be investigated, the mathematical tools of inspection, namely the Raychaudhuri equation and its kinematical components, have to be improved in order to envelop the desired features of the problem. First of all, the congruence is supposed to be uncharged. We therefore can not elicit the physical acceleration from the usual kinematical decomposition. To make use of physical acceleration in this paper, I modify the Raychaudhuri equation, in order to be specified to the gravitational theory in hand. This way, the electromagnetic energy-momentum tensor recasts itself in terms of the gravitational field equations and appears in the kinematical components of a generalized Raychaudhuri equation. As it follows in the next section, this is done through a particular generalization to the base manifold's cotangent bundle. In fact, obtaining generalizations to the Raychaudhuri equation is not something new. There have been numerous publications devoted to this task, each of which, dealing with a peculiar generalization, like $\mathcal{N}$-dimensional spacetimes \cite{Kar1996}, the Arnowitt-Deser-Misner (ADM) formalism \cite{Abreu2011}, generalized theories of gravity \cite{Harko2012}, large-scale bulk motions \cite{Tsagas2013}, bohmian quantum mechanical effects \cite{Das2014} and non-normalized congruences \cite{Fathi2016}. The generalization introduced in this paper, would also cover non-normalized congruences as well as being applicable of inclusion of alternative gravity field equations. 

The paper is organized as follows: In Sec.~\ref{SecII}, the generalized Raychaudhuri equation is introduced which is followed by homothetic filed specifications in Sec.~\ref{SecIII}. In this section, the equation is also used in $f(R)$ theories of gravity to reveal its capabilities. In Sec.~\ref{SecIV}, homothetic congruences are allowed to evolve in the presence of an electrically charged static black hole and their physical and kinematical characteristics are discussed. I show that how these congruences represent very peculiar fates during their evolution. I conclude in Sec.~\ref{SecV}.

\section{A Generalized Raychaudhuri Equation Associated with the Cotangent Bundle}\label{SecII}

\begin{figure}[htp]
\center{\includegraphics[width=11cm]{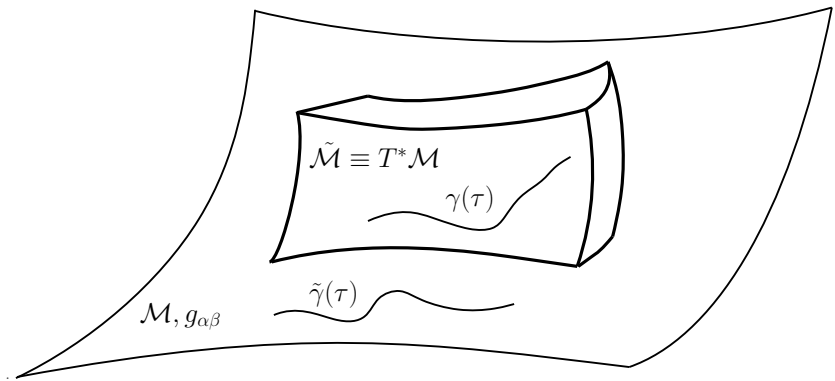}
\caption{\label{fig:1} The cotangent bundle can be regarded as a medium residing in the base manifold. The solution curve $\gamma(\tau)$ is defined on $T^*\mathcal{M}$ whereas the trajectories $\tilde{\gamma}(\tau)$ are measured by the observers who live in $\mathcal{M}$.}}
\end{figure}

The Raychauduri equation is known as governing the evolution of the orthogonal 3-dimensional hypersurface, transverse to the tangential 4-velocity vector, which characterizes the congruence \cite{Poisson2009}. This means that the whole system is defined on the tangent bundle of the base manifold where the Lagrangian mechanics is introduced. In this section, I intend to give a generalization to the Raychaudhuri equation which uses both the tangent and the cotangent bundles of the base manifold to give also the possibility to work with Hamiltonian mechanics. In this regard, the cotangent bundle $T^*\mathcal{M}$ of the base manifold $\mathcal{M}$ (i.e. the Hamiltonian system) is supposed to be an auxiliary medium residing on $\mathcal{M}$ (see Fig.~\ref{fig:1}). Such an approach gives us the opportunity to include the momentum in characterizing the evolution of the congruence. The concept of using the cotangent bundle as a medium, where the 4-momentum originates from, has been first brought and discussed rigorously in Ref. \cite{Thompson2017}.
First of all, let me make some clarifications on the approach. As it has been illustrated schematically in Fig. \ref{fig:1}, the dynamics of a Hamiltonian system is given by the solution curve $\gamma(\tau)$, residing in the cotangent bundle $T^*\mathcal{M}$ of the base manifold $\mathcal{M}$. The associated tangential vector to this curve, is $\bm X = \left(\frac{\partial H}{\partial p_\alpha}, -\frac{\partial H}{\partial x^\alpha}\right)\equiv \left(u^\alpha, \dot p_\alpha\right)$ which lives in the tangent bundle of $T^*\mathcal{M}$. In this case, $(x^\alpha, p_\alpha)$ are the coordinates in $T^*\mathcal{M}$. To find the particles' trajectory on the base manifold, we need $x^\alpha(\tau)$ which is simply obtained by performing an identity map on the $\bu$ sector of $\bm X$ which kills off $\dot{\bp}$. However, to deal with the dynamics of the particles' congruence, we associate their trajectory in $\mathcal{M}$ with the covector field $\bp$ as the whole system's energy/momentum, in $T^*\mathcal{M}$. Since $\bm p$ is a 1-form basis, then one can infer $\ed\bm p = 0$, which in the base manifold gives $p_{\alpha;\beta} = p_{\beta;\alpha}$. 

Having this association, one can construct a projector to the 3-dimensional transverse subspace of the base manifold (3-dimensional space-like foliation, orthogonal to $\bu$) which is characterized by
\begin{equation}\label{eq:projector}
h{_\alpha}^\beta = \delta_\alpha^\beta - (\bm p \cdot\bm u)^{-1} p_\alpha u^\beta,
\end{equation}
satisfying $h{_\alpha}^\beta p_\beta = u^\alpha h{_\alpha}^\beta = \bm 0$, $h{_\lambda}^\alpha h{_\alpha}^\beta = {h_{\lambda}}^\beta$ and ${h_\alpha}^\alpha = 3$. Therefore, we can eliminate any longitudinal components of the tangent bundles of both cotangent bundle and the base manifold, by means of projecting the congruence by the projector given in Eq.~(\ref{eq:projector}). We look for the evolution of this transverse subspace (orthogonal to the world-lines of the observer characterized by $\bm u$), along the congruence (trajectories) in the base manifold associated with its cotangent bundle. This will provide us the Raychaudhuri equation. To proceed with the mathematical procedure, let me introduce the following tensor on $\mathcal{M}$:
\begin{equation}\label{eq:B}
B{^\alpha}_\beta = u{^\alpha}_{;\beta},
\end{equation}  
in which the covariant derivative is calculated with respect to the metric defined on $\mathcal{M}$. The projection of this tensor onto the 3-dimensional orthogonal subspace is designated as
\begin{equation}\label{eq:barB}
\bar{B}{^\alpha}_\beta = h{_\mu}^\alpha u{^\mu}_{;\nu} h{_\beta}^\nu,
\end{equation}  
according to which, the evolution along the congruence generated by $\bm u$, is given by 
\begin{equation}\label{eq:dotbarB}
\frac{D\bar{B}{^\alpha}_\beta}{\ed\tau}\equiv\dot{\bar{B}}{^\alpha}_\beta = \left(
\dot h{_\mu}^\alpha h{_\beta}^\nu + h{_\mu}^\alpha \dot h{_\beta}^\nu
\right) u{^\mu}_{;\nu}
+h{_\mu}^\alpha h{_\beta}^\nu \frac{D}{\ed\tau}\left(u{^\mu}_{;\nu}\right),
\end{equation}
showing that the last term is purely transverse. The Raychaudhuri equation is indeed the evolution of the fractional rate of change of the transverse subspace, i.e. the expansion. In the language I use here, this is
\begin{equation}\label{eq:barB-expansion}
\Theta = h{_\rho}^\mu u{^\rho}_{;\beta} h{_\mu}^\beta \equiv \bar B{^\mu}_\mu,
\end{equation}
whose evolution is given by the following generalized Raychaudhuri equation associated with $T^*\mathcal{M}$ (see appendix \ref{appII}):
\begin{multline}\label{eq:Raychaudhuri-generalized}
\dot\Theta = -\frac{1}{3}\Theta^2 - \sigma_{\mu\nu}\sigma^{\mu\nu} + \omega_{\mu\nu}\omega^{\mu\nu} - R_{\mu\nu} u^\mu u^\nu + a{^\mu}_{;\mu}\\
 - (\bm p\cdot\bm u)^{-1}\left[\dot{(\bm p\cdot \bm a)}
+ 2p_\nu u{^\nu}_{;\rho} a^\rho\right] + (\bm p\cdot\bm u)^{-2}\left[
(\bm p\cdot\bm a)^2
+\dot{(\bm p\cdot \bm u)}(\bm p\cdot \bm a)
\right].
\end{multline} 
The kinematical quantities defined in Eq.~(\ref{eq:Raychaudhuri-generalized}) are 
\begin{subequations}\label{eq:kinematicalQuantities}
\begin{align}
\Theta = u{^\mu}_{;\mu} - (\bm p\cdot \bm u)^{-1}(\bm p\cdot\bm a),\label{eq:kinematicalQuantities-a}\\
\sigma_{\mu\nu} = h{_\mu}^\alpha u_{(\alpha;\beta)} h{_\nu}^\beta - \frac{1}{3}\Theta h_{\mu\nu},\label{eq:kinematicalQuantities-b}\\
\omega_{\mu\nu} = h{_\mu}^\alpha u_{[\alpha;\beta]} h{_\nu}^\beta,\label{eq:kinematicalQuantities-c}
\end{align}
\end{subequations}
which are respectively, the expansion scalar, the symmetric shear and the anti-symmetric vorticity tensors. Also, $h_{\mu\nu}= h{_\mu}^\alpha g_{\alpha\nu}$. Note that, whether or not $\bm p\cdot\bm u = \mathrm{const.}$, the expansion in Eq.~(\ref{eq:kinematicalQuantities-a}) retains the common form of $\Theta = u{^\mu}_{;\mu}$, iff the congruence is geodesic ($\bm a =\bm 0$). I should note here that the existence of the $\bp$ covector in the generalized Raychaudhuri equation, aims at providing an association to the energy-momentum resources of the gravitational theory in hand. This means that, this covector associates the congruences' evolution to the configuration of the gravitational source, described within a particular theory of gravity. In this regard, it is different from the usual conception of the momentum in general relativity where we put $\bp = m\bu$ ($m=\mathrm{const.}$) and $\bu$ is considered normalized. For this latter case, the above generalized Raychaudhuri equation will reduce to its usual textbook version (see appendix \ref{appII})
\begin{eqnarray}\label{eq:Raychaudhuri-common}
\dot\Theta &=& -\frac{1}{3}\Theta^2 - \sigma_{\mu\nu}\sigma^{\mu\nu} + \omega_{\mu\nu}\omega^{\mu\nu} - R_{\mu\nu} u^\mu u^\nu + a{^\mu}_{;\mu}.
\end{eqnarray}
In this case, the generalized expansion in Eq.~(\ref{eq:kinematicalQuantities-a}) reduces to $\Theta = u{^\mu}_{;\mu}$, whether or not the congruence is geodesic, because in this special case we have $\bm u\cdot\bm a = \frac{1}{2}(u^\mu u_\mu)_{;\nu} u^\nu = 0$. The generalized equation however, helps us to specify the particles' motion to the cotangent bundle, where the Hamiltonian mechanics is defined. In the current study, the cotangent bundle exhibits the characteristics of the energy-momentum tensor, by exploiting its geometrical correspondent, namely the gravitational field equations. In this regard, we can perform investigations in different theories of gravity, by specifying the transverse subspace to those theories. When an energy-momentum tensor $T_{\alpha\beta}$ can be defined within a theory of gravity, the momentum density covector and the energy are defined as
\begin{subequations}\label{eq:momentum-energy-T}
\begin{align}
p_\alpha \doteq -T_{\alpha\beta} u^\beta,\label{eq:momentum-energy-T-a}\\
E \doteq -\bm p\cdot\bm u =  T_{\alpha\beta} u^\alpha u^\beta.\label{eq:momentum-energy-T-b}
\end{align}
\end{subequations}
Accordingly, the term $2p_\nu u{^\nu}_{;\rho}a^\rho$ in Eq.~(\ref{eq:Raychaudhuri-generalized}), could be expanded as
\begin{eqnarray}\label{eq:theTerm2pua}
2p_\nu u{^\nu}_{;\rho}a^\rho &=& 2 \left[(p_\nu u^\nu)_{;\rho} - p_{\nu;\rho} u^\nu\right] a^\rho = -2\left[E_{,\rho} + p_{\rho;\nu}u^\nu\right]a^\rho\nonumber\\
&=&-2\left(E_{,\rho}+\dot p_\rho\right)a^\rho,
\end{eqnarray}
where I have used the fact that $\ed\bm p = 0$. To proceed further, I apply the above results and definitions, to show that how the above generalized components can be used in a way to include alternative theories of gravity. In the next section, the generalized expansion is dealt with in the context of homothetic congruences and some physical properties are discussed.

\section{ The Generalized Congruence Evolution in $f(R)$-Gravity
}\label{SecIII}

I start with an extended theory of gravity, namely $f(R)$-gravity, to demonstrate the abilities of the generalized kinematical components of the Raychaudhuri equation, to be specified to a particular theory of gravity. The $f(R)$-gravity theory is the most natural alternation to general relativity,  relevant to the dark energy scenario, and has received a great deal of attention from the community. Since the modification is based on the Ricci scalar, the reduction to general relativity (which is the goal of this paper) is therefore trivial. The $f(R)$ theories of gravity, are approached by means of the generalized gravitational action (in the $G = c =1$ units)\cite{Sotiriou2010}
\begin{equation}\label{eq:VII}
\mathcal{S}_{f(R)}=\frac{1}{\kappa }\int \ed^4x \sqrt{-g}~ f(R) +\int \ed^4x L_m\left(g_{\alpha \beta },\Psi _m\right),
\end{equation}
where $L_m$ is the matter Lagrangian density and $g=\mathrm{det}(g_{\alpha\beta})$. The least action condition, imposed on the variation of the above relation in terms of $g^{\alpha\beta}$, (that is $\frac{\delta \mathcal{S}_{f(R)}}{\delta g^{\alpha \beta }}=0$), results in the following field equations: 
\begin{equation}\label{eq:VIII}
-\frac{1}{2} f(R) g_{\mu \nu }+g_{\mu \nu } \square F(R)-\nabla _{\mu }\nabla _{\nu }F(R)+F(R) R_{\mu \nu }=\kappa  T_{\mu \nu },
\end{equation}
in which $\kappa=8 \pi$ and $F =f_{,R}$. Moreover, the energy-momentum tensor is defined as
\begin{equation}\label{eq:IX}
T_{\mu \nu }=-\frac{2}{\sqrt{-g}}\frac{ \delta\left(  \sqrt{-g} L_m\right)}{ \delta g^{\mu \nu }}.
\end{equation}
To specify the kinematical characteristics of time-like (or space-like) congruences in the context of the Raychaudhuri equation in Eq. (\ref{eq:Raychaudhuri-generalized}) and in the $f(R)$-gravity realm, one needs to write down the momentum and energy in Eqs. (\ref{eq:momentum-energy-T}), based on the field equations (\ref{eq:VIII}). Accordingly, one obtains
\begin{subequations}\label{eq:X}
\begin{align}
p_{\mu }=-\frac{1}{\kappa }\left(F R_{\mu \nu } u^{\nu }-\frac{1}{2} f g_{\mu \nu } u^{\nu }+g_{\mu \nu }F^{;\lambda }{}_{;\lambda }u^{\nu }-F_{;\mu \nu }u^{\nu }\right),\label{eq:Xa}\\
E=\frac{1}{\kappa }\left(F R_{\mu \nu } u^{\mu } u^{\nu }-\frac{1}{2}f(\bu\cdot\bu)+F^{;\lambda }{}_{;\lambda }(\bu\cdot\bu)-F_{;\mu \nu }u^{\mu }u^{\nu }\right),\label{eq:Xb}
\end{align}
\end{subequations}
in which for convenience, I have dropped mentioning the $R$-dependence. In the above relations, the momentum and energy have been given completely geometrical definitions. Note that, like the momentum, the energy is also provided by associating the congruence with the cotangent bundle. This energy would therefore constitute the congruence energy. This is the key to the interpretation of our generalized Raychaudhuri equation in any particular theory of gravity. Now to talk about congruence evolution, I turn to the kinematical parameters in $f(R)$-gravity. Most importantly, one notices the presence of the acceleration term in the important expansion relation in Eq. (\ref{eq:kinematicalQuantities-a}). As mentioned before, I am are interested in homothetic congruences in this study. I therefore, use the relation in Eq. (\ref{eq:III}), together with Eq. (\ref{eq:kinematicalQuantities-a}), to get
\begin{multline}\label{eq:XI}
\Theta =u^{\alpha }{}_{;\alpha }-\left[F R_{\mu \nu } u^{\mu } u^{\nu }-\frac{1}{2}f(\bu\cdot\bu)+F^{;\alpha }{}_{;\alpha }(\bu\cdot\bu)-F_{;\mu \nu }u^{\mu }u^{\nu }\right]^{-1}\\
\times\left\{\dot{(\bu\cdot\bu)}\left(\frac{1}{4}f-\frac{1}{2}F^{;\alpha }{}_{;\alpha }\right)+\xi (\bu\cdot \bu)\left(-f+2F^{;\alpha }{}_{;\alpha }\right)\right.\\
\left.+2\xi \left(F R_{\mu \nu }-F_{;\mu \nu }\right)u^{\mu }u^{\nu }+\frac{1}{2}\left(g^{\lambda \mu }F_{;\mu \nu }-F R^{\lambda }{}_{\nu }\right)(\bu\cdot\bu)_{;\lambda}u^{\nu }\right\}.
\end{multline}
The expansion scalar, is the outstanding characteristic of the congruence evolution because it talks about the fractional rate of change in terms of the congruence parameterization. This gives us insight into the shape and the intensity of the the energy flow. The fractional rate of change given in Eq.~(\ref{eq:XI}) however, belongs to that of homothetic congruences in $f(R)$-gravity. I should here note that, although a $f(R)$-generalization of Raychaudhuri equation and its kinematical components has been provided in Ref.~\cite{Harko2012}, however, the approach used in that paper is crucially based on the energy-momentum of a perfect fluid whose evolution is essentially geodesic and normalized (because it is comoving with the spacetime). Here, to obtain the generalized Raychaudhuri equation, I basically modify the transverse subspace in order to have conformity with all kind of congruences within a desired theory of gravity. The generalization therefore, is not confined to any particular source of energy.

It can be easily seen that for normalized congruences (i.e. $\bu\cdot\bu=\mathrm{const.}$), the above expansion reduces to 
\begin{equation}\label{eq:XII}
\Theta_{\mathrm{norm}} =u^{\alpha }{}_{;\alpha }-2 \xi,
\end{equation}
which is the standard textbook relation for non-affinely-parameterized congruences. This shows that the generalized Raychaudhuri equation in Eq. (\ref{eq:Raychaudhuri-generalized}), is consistent with the usual form, once the standard premises are satisfied. Furthermore, the generalized kinematical parameters are capable of specifying the congruence evolution, for particular field equations. For this reason, I will take care with non-normalized homothetic congruences, later in this study. But at this stage, let me rely on a spherically symmetric spacetime manifold, described by the metric   
\begin{equation}\label{eq:XIII}
\ed s^2 = -A(r)\ed t^2 + B(r)^{-1} \ed r^2 + r^2 \ed \theta^2 + r^2\sin\theta^2\ed\phi^2,
\end{equation}
in the $(t ,r ,\theta, \phi)$ chart. Now to obtain the radially infalling homothetic fields, as the generators of the congruences in the above spacetime, I adopt the generator to be of the form
\begin{equation}\label{eq:XIV}
u^\alpha = \left(u^0(r), u^1(r), 0, 0\right).
\end{equation}
The above field is homothetic, if it obviates the condition in Eq. (\ref{eq:I}). Note that, for the above vector field to be future-directed, one suitable choice for infalling observers is $u^0>0$ and $u^1<0$. The congruence generator in Eq. (\ref{eq:XIV}) leads to four independent equations, solving which, we obtain three non-normalized generators as
\begin{subequations}\label{eq:XV}
\begin{align}
u_1^{\alpha }=\left(a, \frac{2\xi A  }{A'}, 0, 0\right),\label{eq:XVa}\\
u_2^{\alpha }=\left(a, b\sqrt{B} + \sqrt{B} \int{\frac{\xi}{\sqrt{B}}\ed r}, 0, 0\right),
\label{eq:XVb}\\
u_3^\alpha = \left(a, \xi r, 0 ,0\right),
\label{eq:XVc}
\end{align}
\end{subequations}
in which, $a$ and $b$ are integration constants, the prime stands for differentiation with respect to $r$ and I have dropped mentioning the $r$-dependence. Note that, the complete identification of $\bu_2$ depends directly on the specified solutions for the metric (\ref{eq:XIII}). The above vector fields are essentially independent and give different values for $\bu\cdot\bu$. In fact we have
\begin{subequations}\label{eq:XVI}
\begin{align}
\bu_1\cdot \bu_1=A \left(\frac{4 \xi ^2 A }{B \left(A'\right)^2}-a^2\right),\label{eq:XVIa}\\
\bu_2\cdot \bu_2=\left(b +  \int{\frac{\xi}{\sqrt{B}}\ed r}\right)^2-a^2 A,
\label{eq:XVIb}\\
\bu_3\cdot\bu_3 = \frac{\xi ^2 r^2}{B}-a^2 A.
\label{eq:XVIc}
\end{align}
\end{subequations}
Once the metric potential is specified to a particular $f(R)$ theory of gravity, these values can be applied in the generalized expansion relation, to inspect the cross-sectional rate if change of the homothetic congruences in the context of that theory. These values also highlight the importance of the existence of $\bu\cdot\bu$ terms in the generalized kinematical formulation, introduced above. Since the congruence generators are no longer normalized, it is indispensable for the expansion relation to contain the values obtained in Eqs. (\ref{eq:XVI}), to enable us talking about the congruence evolution, peculiar to our chosen theory.

\subsection{Congruence Intensity}

\begin{figure}[t]
\center{\includegraphics[width=11cm]{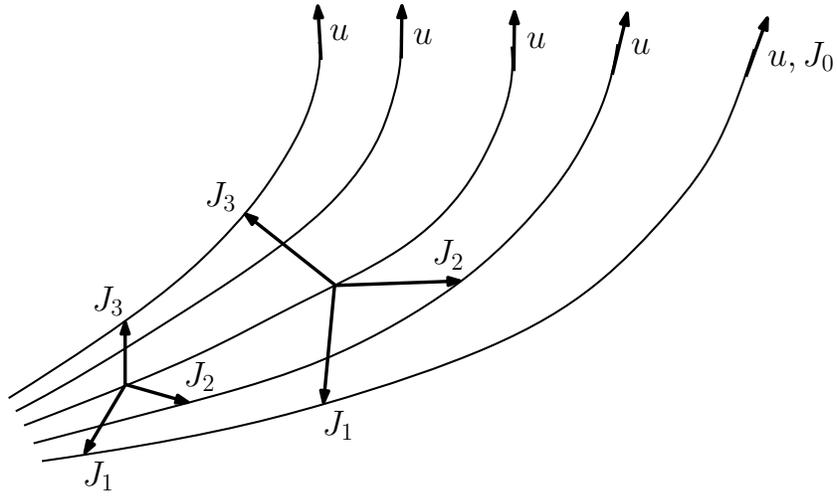}
\caption{\label{fig:2}  A congruent flow generated by the tangential vector $\bu$. There are a set of Jacobi fields $\{\bJ_A\}$ which measure the deformation of the flow. 
}}
\end{figure}

There is one interesting concept which is worth discussing in this section; the congruence intensity. In classical optics, the field intensity is defined as the light power passing a definite surface. This is indeed the way an observer detects the light and measures its saturation. In the context of infalling congruences however, one is supposed to take account of the observer-dependence property of the fractional rate of change and the volume filled by the congruences in 4-dimensions. I therefore define the congruence power as
\begin{equation}\label{eq:XVII}
P=\frac{\ed E}{\ed \tau }=E_{,\mu }u^{\mu },
\end{equation}
which in this study, the energy $E$ has been defined in Eq. (\ref{eq:Xb}). On the other hand, the 4-dimensional volume (4-volume filled by the congruence), is described in terms of a set of Jacobi fields $\{\bJ_A\}$, $A=0,1,2,3$. We have \cite{Frankel2012}
\begin{equation}\label{eq:XVIII}
V_{\Omega }=\Omega _{\alpha \beta \mu \nu } J_0^{\alpha } J_1^{\beta } J_2^{\mu } J_3^{\nu },
\end{equation}
with
\begin{equation}\label{eq:XIX}
\Omega _{\alpha \beta \mu \nu }=\sqrt{-g} \epsilon _{\alpha \beta \mu \nu }
\end{equation}
to be the volume 4-form where $\epsilon_{\alpha\beta\mu\nu}$ is the Levi-Civita symbol. Indeed, $\{\bJ_A\}$ can be regarded as the legs of a 4-dimensional parallelepiped, the volume of which is $V_\Omega$. As it is seen in Fig.~\ref{fig:2}, the Jacobi fields govern the deformation of the congruence along with its flow. 
I therefore write the congruence intensity as follows:
\begin{equation}\label{eq:XX}
I=\frac{P}{V_{\Omega }}.
\end{equation}
Now let me find this relation for the spacetime given in Eq. (\ref{eq:XIII}), for those congruences provided in Eqs. (\ref{eq:XV}). One essential condition for Jacobi fields, obtained form Eq. (\ref{eq:IV}), is that the Jacobi fields are Lie transported along the congruence, i.e. $\mathcal{L}_{\bu}\bJ_A = \mathbf{0}$. Therefore, the most natural choice could be $\bJ_0\equiv\bu$ (because every vector field is Lie transported along itself). So, based on the fields in Eqs. (\ref{eq:XV}), I let
\begin{equation}\label{eq:XXI}
^1\bJ_0 = \bu_1,\quad\quad ^2\bJ_0 = \bu_2,\quad\quad ^3\bJ_0 = \bu_3,\quad\quad
\end{equation}
in which the superscripts on the left side of the Jacobi fields, indicate the relevance to individual congruence generators. To proceed with obtaining the other vectors, let me put
\begin{equation}\label{eq:XXII}
J_1^{\alpha }=\left(0,J_1^1(r),0,0\right).
\end{equation}
Applying the Lie transport condition for each of the vector fields in Eqs. (\ref{eq:XV}), it is determined that
\begin{subequations}\label{eq:XXIII}
\begin{align}
^1J_1^\alpha = \left(0, \frac{A}{A'}, 0, 0\right),\label{eq:XXIIIa}\\
^2J_1^\alpha = \left(0, b\sqrt{B} + \sqrt{B} \int{\frac{\xi}{\sqrt{B}}\ed r}, 0, 0\right),
\label{eq:XXIIIb}\\
^3J_1^\alpha = \left(0, r, 0, 0\right).
\label{eq:XXIIIc}
\end{align}
\end{subequations}
To retain orthogonality and Lie transportation, the simplest choices for the other two fields will be
\begin{subequations}\label{eq:XXIV}
\begin{align}
^1J_2^\alpha = ^2J_2^\alpha = ^3J_2^\alpha = \left(0, 0, 1, 0\right),\label{eq:XXIVa}\\
^1J_3^\alpha = ^2J_3^\alpha = ^3J_3^\alpha = \left(0, 0, 0, 1\right).\label{eq:XXIVb}
\end{align}
\end{subequations}
Now that the Jacobi fields have been specified, I apply Eq. (\ref{eq:XVIII}), to obtain the congruences' 4-volumes. In the equatorial plane ($\theta=\frac{\pi}{2}$) these are 
\begin{subequations}\label{eq:XXV}
\begin{align}
^1V_{\Omega }=\frac{a r^2 A^{3/2} }{\sqrt{B} A'},\label{eq:XXVa}\\
^2V_{\Omega }=a r^2 \sqrt{A} \left(b + \int{\frac{\xi}{\sqrt{B}}\ed r}\right),\label{eq:XXVb}\\
^3V_{\Omega }=\frac{a r^3 \sqrt{A} }{\sqrt{B}}.\label{eq:XXVc}
\end{align}
\end{subequations}
One can then obtain three separate relations for the congruence intensities (i.e. $^1I$, $^2I$ and $^3I$), by employing Eqs. (\ref{eq:Xb}), (\ref{eq:XV}) and (\ref{eq:XXV}). To get sensible results, this however needs a complete specification of the theory, which is what I deal with in the next section.

\section{The Homothetic Congruences in the Gravitational Field of a Charged Massive Source}\label{SecIV}

The approach is generally based on a homothetic congruence of uncharged particles, who fall onto a static charged black hole of mass $M$ and charge $Q$. This way, a non-vacuum solution of the gravitational field equations is needed. The reason of adoption of a charged source, is mostly based on generating the simplest non-vacuum gravitational field, to examine the congruence expansion with non-zero energy. The electromagnetic energy-momentum tensor 
\begin{equation}\label{eq:XXVI}
T_{\mu \nu }=\frac{1}{4\pi} \left(F_{\mu \alpha } F^{\alpha }{}_{\nu }-\frac{1}{4} g_{\mu \nu } F^{\alpha \beta } F_{\alpha \beta } \right),
\end{equation}
with $F_{\alpha\beta} = \mathcal{A}_{\beta;\alpha} - \mathcal{A}_{\alpha;\beta}$ as the field strength tensor, is specified by the 4-potential $\mathcal{A}^\alpha = \left(\Phi(r), 0, 0, 0\right)$ for static spherically symmetric sources of charge $Q$, with $\Phi(r)\sim\frac{1}{4 \pi}\frac{Q}{r}$ to be the electric potential due to such sources. One recalls the existence of a natural acceleration $a^\mathrm{EM}_\mu = \frac{q}{m} F_{\mu\nu} u^\nu$ for charged particles of mass $m$ and charge $q$ who fall in the above electromagnetic field \cite{MTW1973}.  However since my approach is based on uncharged particles, this acceleration no longer exists. The congruence however, experiences a geometric acceleration due to its homothetic nature. My aim is to examine the effects of the spacetime generated by a static charged black hole on the aforementioned congruence by means of the generalized values in Eqs. (\ref{eq:X}) and (\ref{eq:XI}). These generalized components, because of their dependence on the electromagnetic energy-momentum tensor, would therefore show the effect of the black hole's charge on the congruence, despite its electrical neutrality. 

Now since the whole investigation stems in the spacetime geometry, the gravitational field equations should therefore be solved for this source. The resultant spacetime geometry is exploited to provide expressions for the dynamical and kinematical characteristics of infalling congruences. Note that, since I have generically replaced the energy-momentum tensor with its field equation correspondent (what I did in obtaining Eqs. (\ref{eq:X})), instead of relying on the above electromagnetic energy-momentum tensor, the only thing we need is the solution to the field equations for a particular charged source, in the context of the chosen theory of gravity.

\subsection{The General Relativistic Limit}

\begin{figure}[t]
\center{\includegraphics[width=8.5cm]{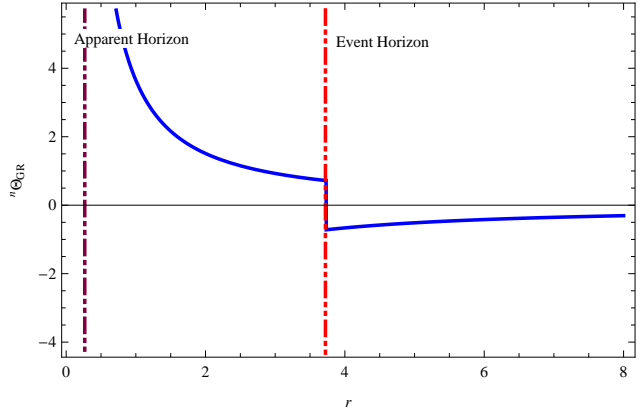}
\caption{\label{fig:2'} The near-horizon expansion of a normalized congruence of charged particles in the RN spacetime for values $M=2$, $Q=1$, $m=0.01$, $q=0.005$ and $c_1=1.5$}}
\end{figure}

To provide insights into the approach, the best choice is to work with the most essential theory of gravity, namely general relativity. General relativity, is indeed an $f(R)$ theory of gravity which designates $f(R)=R$ in Eq. (\ref{eq:VII}). In general relativity, the spacetime geometry describing a static charged massive source, is the Reissner-Nordstr\"{o}m (RN) solution, which lets
\begin{equation}\label{eq:XXVII}
A(r) = B(r) = 1-\frac{2 M}{r}+\frac{Q^2}{r^2},
\end{equation}
for the metric (\ref{eq:XIII}). The above value vanishes for $r_\pm = M \pm \sqrt{M^2-Q^2}$ being the locations for the event (outer) and apparent (inner) horizons \cite{Poisson2009}. In the above metric potentials, $M$ is the Schwarzschild massive source and in the additional terms to Schwarzschild metric, $Q$ is the charge factor\footnote{In the geometric units I use here, i.e. $G=c=1$, it is $\mathrm{dim}[M] = \mathrm{m}$, $\mathrm{dim}[Q] = \mathrm{m}$ and $\mathrm{dim}[\xi] = \mathrm{s}^{-1}$.}. Motion of charged particles around RN black-holes has been discussed vastly in the literature (e.g. in Ref. \cite{Chandrasekhar1998} for general discussion and in Ref. \cite{Fathi2013} for the case of relativistic stars). In the context of congruences however, some more inspections are needed, since particle paths are now regarded as a set, and therefore some more kinematical characteristics now take part in the process. To demonstrate the characteristics of a congruence of falling charged particles in the RN geometry, let me consider a vector field 
\begin{equation}\label{eq:RN-1}
v^\alpha = \left(v^0(r),v^1(r),0,0\right),
\end{equation}
as the congruence generator. As mentioned before, since the above vector field is the congruence generator of falling test particles of mass $m$ and charge $q$ in an electromagnetic field, the physical acceleration equation
\begin{equation}\label{eq:RN-1-1}
v{^\alpha}_{;\beta}v^\beta = a^\alpha_{\mathrm{EM}} = \frac{q}{m} F{^\alpha}_\beta v^\beta,
\end{equation}
should be satisfied. Applying Eq. (\ref{eq:RN-1}) to Eq. (\ref{eq:RN-1-1}), the only possible choice is obtained as
\begin{equation}\label{eq:RN-2}
v^\alpha = \left(
\frac{c_1 r^2}{\chi }-\frac{q Q}{4 \pi  m r},-\frac{\sqrt{\chi }}{r^2} \sqrt{\chi  \left(\frac{c_1 r^2}{\chi }-\frac{q Q}{4 \pi  m r}\right)^2-r^2},0,0
\right),
\end{equation}
for a future directed normalized congruence, in which $\chi=Q^2-2 M r + r^2$ is a scalar of dimensions of $\mathrm{m}^2$ and $c_1$ is a constant. Note that, the above congruence is normalized and indeed $\bm v\cdot\bm v=-1$, which is a common condition for moving particles in gravitational fields. To see the congruence evolution in a near-horizon region, it is of benefit to calculate the congruence expansion 
The congruence generator in Eq. (\ref{eq:RN-2}) is not homothetic, however it undergoes an acceleration given in Eq. (\ref{eq:RN-1-1}). We therefore should apply the original generalized relation in Eq.~(\ref{eq:kinematicalQuantities-a}) for the expansion. Note that, since $\bm v$ is normalized, one can infer that $v_\alpha v{^\alpha}_{;\beta}v^\beta \equiv \bm v\cdot\bm{a}_{\mathrm{EM}} = 0$. On the other hand, to keep the conformity with the current study, the association to the cotangent bundle is also retained. We therefore possess a congruence of non-geodesic charged particles, for which, applying Eqs.~(\ref{eq:X}) in the general relativistic limit, yields
\begin{equation}\label{eq:RN-2-1}
^n\Theta_{\mathrm{GR}} = v{^\sigma}_{;\sigma} - \frac{q}{m}\frac{R_{\mu\nu}F{^\mu}_\lambda v^\lambda v^\nu}{R_{\alpha\beta} v^\alpha v^\beta + \frac{1}{2}R},
\end{equation}
in which the superscript $n$ in the left, indicates the normalized property and the GR subscript stands for the general relativistic limit, which is being used in this section. Using this relation for the normalized congruence given in Eq.~(\ref{eq:RN-2}), we get 
\begin{equation}\label{eq:RN-3}
^n\Theta _{\text{GR}}=\frac{16 \pi ^2 m^2 r^4 \left(\chi-2 c_1^2 r^2-M r+r^2 \right)+4 \pi  c_1 m q Q r^3 \left(\chi-2 M r+2 r^2 \right)+q^2 Q^2 \chi  \left(Q^2-r^2\right)}{4 \pi  m r^4 \sqrt{16 \pi ^2 m^2 r^4 \left(c_1^2 r^2-\chi \right)-8 \pi  c_1 m q Q r^3 \chi +q^2 Q^2 \chi ^2}}.
\end{equation}
The behavior of the above expansion, has been shown in Fig. \ref{fig:2'}. 
As it is seen, outside the event horizon, the congruence experiences a steady convergence, until it falls onto the horizon. Note that, the congruence is not completely focused on the horizon, so that the shift from being convergent to begin divergent, becomes possible. This is what happens for the congruence by passing the event horizon. As shown in the figure, the negative expansion becomes positive inside the event horizon until it fully diverges right before the apparent horizon. The outside convergence is an expected feature of infalling congruences as demanded by attractive gravity of black holes. In the special case discussed above, the event horizon is also responsible for the sign changing of the expansion. However, as it will be shown in what follows, the above feature do alter remarkably for the case of homothetic congruences. The homothetic congruences in this study are supposed to be essentially uncharged and possess natural geometric acceleration. We therefore can expect different results from those of charged particles congruence discussed above. In what I will illustrate in this section, I show that the congruences' fates are different from what are expected to be and indeed have some similarities with {\textit{particles' trajectories}}. In this section, I will present some of these results and discuss them.\\

Now let me determine the homothetic congruence generators in Eqs. (\ref{eq:XV}) for this spacetime, which will be dealt with, individually. Note that, all the three congruences are non-normalized, which highlights the importance of the existence of the $\bu\cdot\bu$ term in the generalized Raychaudhuri equation and its kinematical components.

\subsubsection{The $u_1$ Congruence}

\begin{figure}[t]
\center{\includegraphics[width=8.5cm]{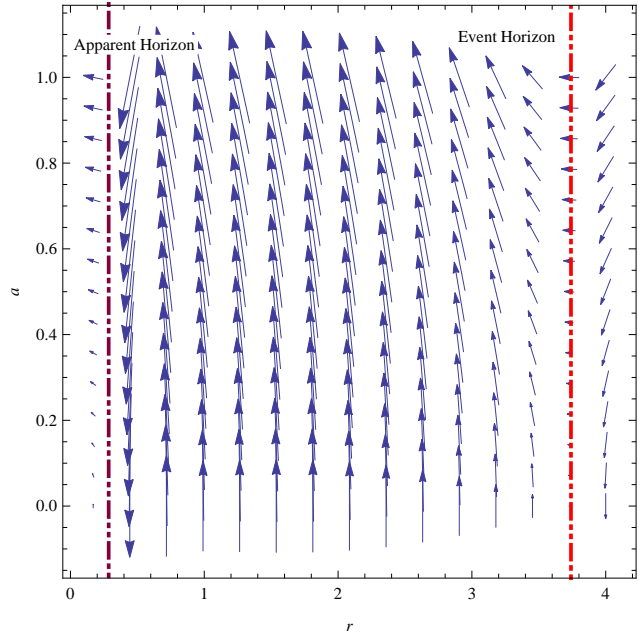}
\caption{\label{fig:4} The $\bu_1$ vector field which indicates the congruence convergence/divergence while it passes the horizons.}}
\end{figure}

From Eq. (\ref{eq:XVa}), one gets
\begin{equation}\label{eq:XXVIII}
u_1^{\alpha }=\left(a,\frac{\xi  r \chi}{\psi},0,0\right),
\end{equation}
in which $\psi = \left(M r-Q^2\right)$ is a scalar of dimensions of $\mathrm{m}^2$. Equations (\ref{eq:X}) and (\ref{eq:XI}), provide us with the congruence energy and the expansion as
\begin{equation}\label{eq:XXIX}
^1E_{\text{GR}} = \frac{Q^2 \chi \left(a^2 \psi^2-\xi ^2 r^6\right)}{\kappa  r^6 \psi^2},
\end{equation}
\begin{multline}\label{eq:XXX}
^1\Theta _{\text{GR}}=\frac{\xi }{\psi ^2 \left(a^2 \psi ^2-\xi ^2 r^6\right)}
\times
\left[\xi ^2 r^6 \left(4 Q^2 r (2 r-5 M)+M r^2 (11 M-6 r)+7 Q^4\right)-a^2 \psi ^2 \left(Q^2 r (5 r-12 M)\right.\right.\\
\left.\left.+M r^2 (7 M-4 r)+4 Q^4\right)\right].
\end{multline}
Moreover, by applying Eqs. (\ref{eq:XXVa}) for the $\bu_1$ congruence of the energy in Eq. (\ref{eq:XXIX}), one gets
\begin{equation}\label{eq:XXXI}
^1I_{\mathrm{GR}} =  \frac{4 \xi Q^2 \left[
\xi^2\left(Q^2-M^2\right)r^8 - a^2\psi^3\left(3 Q^2+r(2 r - 5 M)\right)
\right]}{a \kappa r^9 \psi^3}.
\end{equation}
To visualize the evolution, I plot the generalized energy $^1E_{\mathrm{GR}}$ and expansion $^1\Theta_{\mathrm{GR}}$. 
\begin{figure}[t]
\center{\includegraphics[width=8.3cm]{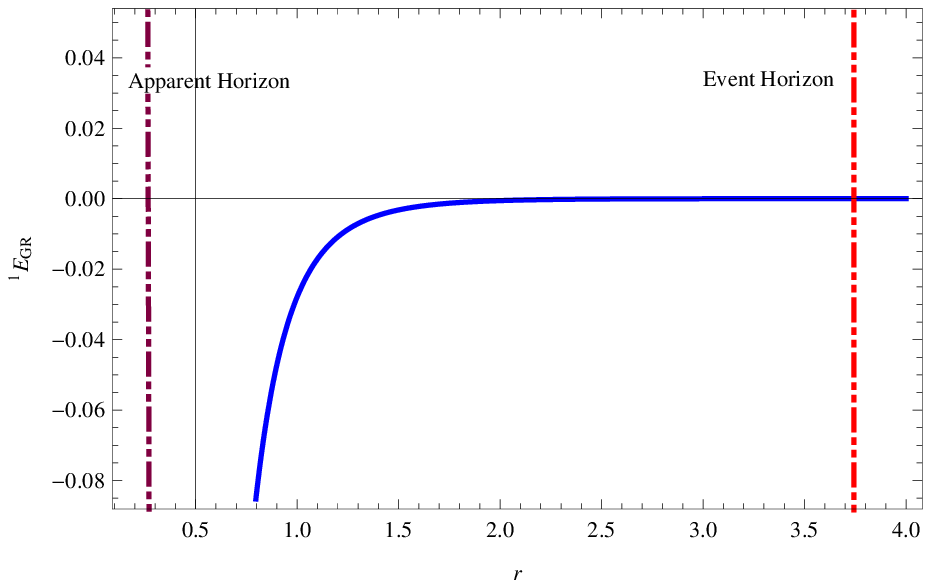}~(a)
\hfil
\includegraphics[width=8.3cm]{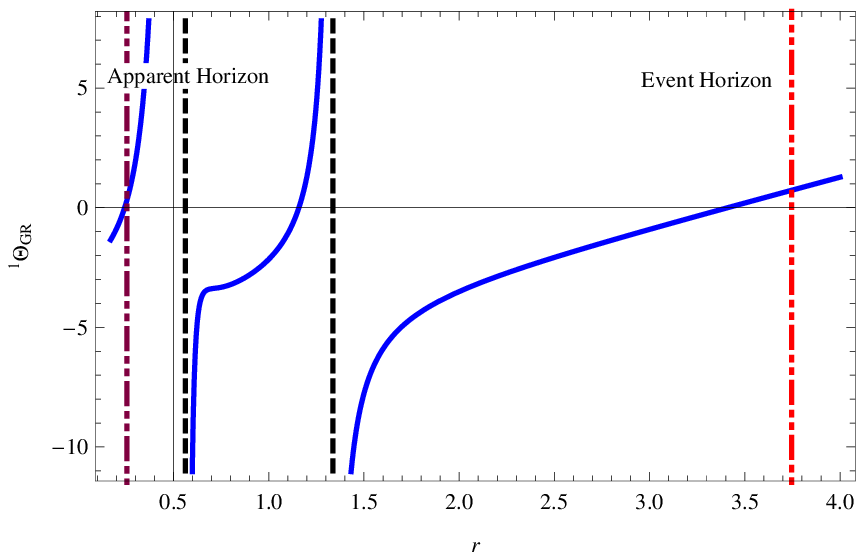}~(b)
\caption{\label{fig:3} The congruence characteristics for $\bu_1$. The figures show (a) the energy and (b) the expansion for the values $a = 0.6$, $M=2$,  $Q=1$ and $\xi=0.1$.}}
\end{figure}
As it can be seen in Fig.~\ref{fig:3}{\color{Blue}{a}}, the congruence energy outside the event horizon is almost zero. This indicates an approximate balance between the positive ({\textit{due to matter}}) and negative ({\textit{due to gravity}}) energies. The congruence will therefore travel freely outside the horizon and because of its essential inward direction, it enters the horizon. Note that, the congruence is initially divergent, showing a remarkable difference o the behavior of charged particle congruence, as I discussed earlier. This divergence however, is reduced under a very slight increase in the gravitational energy. Hence, the congruence finally shifts from being divergent to being convergent inside the event horizon.  At certain points where the gravitational energy experiences a notable increase, the expansion has a singularity, causing the congruence to be completely focused. One therefore concludes that the first congruence focusing occurs inside the black hole region. The focusing point is a singularity in the expansion indicating a cusp within congruence propagation. Consequently, if the congruence is able to travel further, it continues in a dispersed\footnote{The {\textit{dispersion}} term which I use in this discussion, is the exact opposite of the {\textit{focus}} term. In fact, for positive infinite expansion, the congruence would diverge until it looses its integrity (dispersion). In contrast, for negative infinite expansion, the congruence would converge until all of its components are concentrated in a cusp (focusing).} form, as it is shown in Fig.~\ref{fig:3}{\color{Blue}{b}}. This is followed by a second focusing and dispersion around the second expansion singularity. This corresponds to a remarkable increase in the gravitational energy right before the congruence reaches the apparent horizon. This will cause the congruence to pass the apparent horizon in a convergent manner. From a different viewpoint and regarding $\bu_1$ as a vector field, we can see its evolution for different values of $a$ in Fig.~\ref{fig:4}. One can note the shifts in directions around the horizons. Furthermore, the congruence intensity for the $\bu_1$ field has been shown in Fig.~\ref{fig:5} which indicates a raise in the positive ({\textit{due to the attraction within the gravitational energy}}) values, once the congruence energy becomes completely gravitational and large. This situation happens around the apparent horizon.  
\begin{figure}[t]
\center{\includegraphics[width=8.5cm]{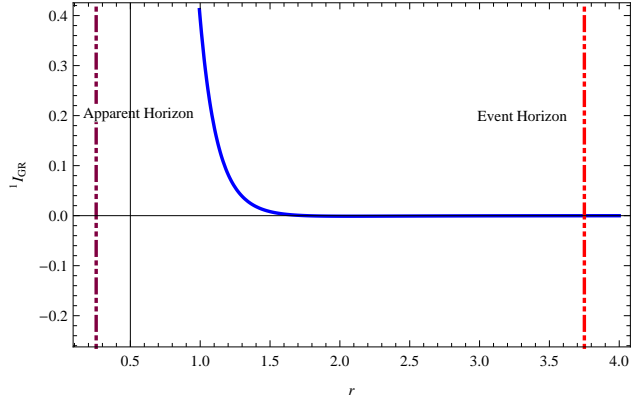}
\caption{\label{fig:5} The congruence intensity $^1I_{\mathrm{GR}}$ for positive energies given in Fig. \ref{fig:3}.}}
\end{figure}

\subsubsection{The $u_2$ Congruence}

\begin{figure}[t]
\center{\includegraphics[width=8.3cm]{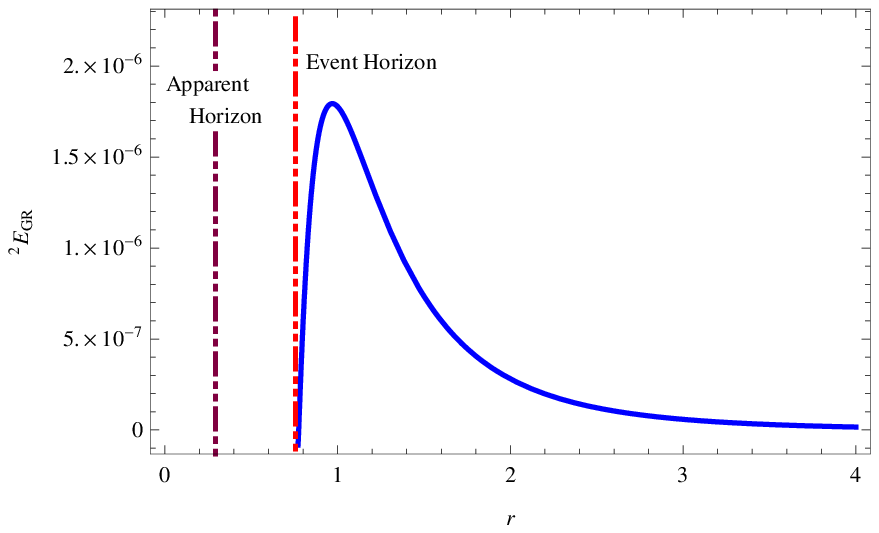}~(a)
\hfil
\includegraphics[width=8.3cm]{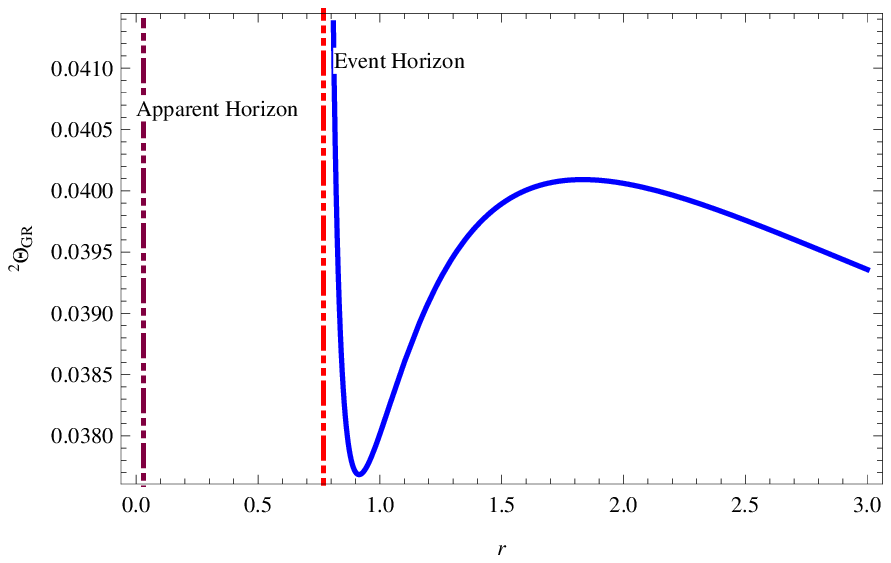}~(b)
\caption{\label{fig:6} The congruence characteristics for $\bu_2$. The figures show (a) the energy and (b) the expansion for the values $a = 0.1$, $b=0.01$, $M=0.4$,  $Q=0.15$ and $\xi=0.011$.}}
\end{figure}

\begin{figure}[t]
\center{\includegraphics[width=8.5cm]{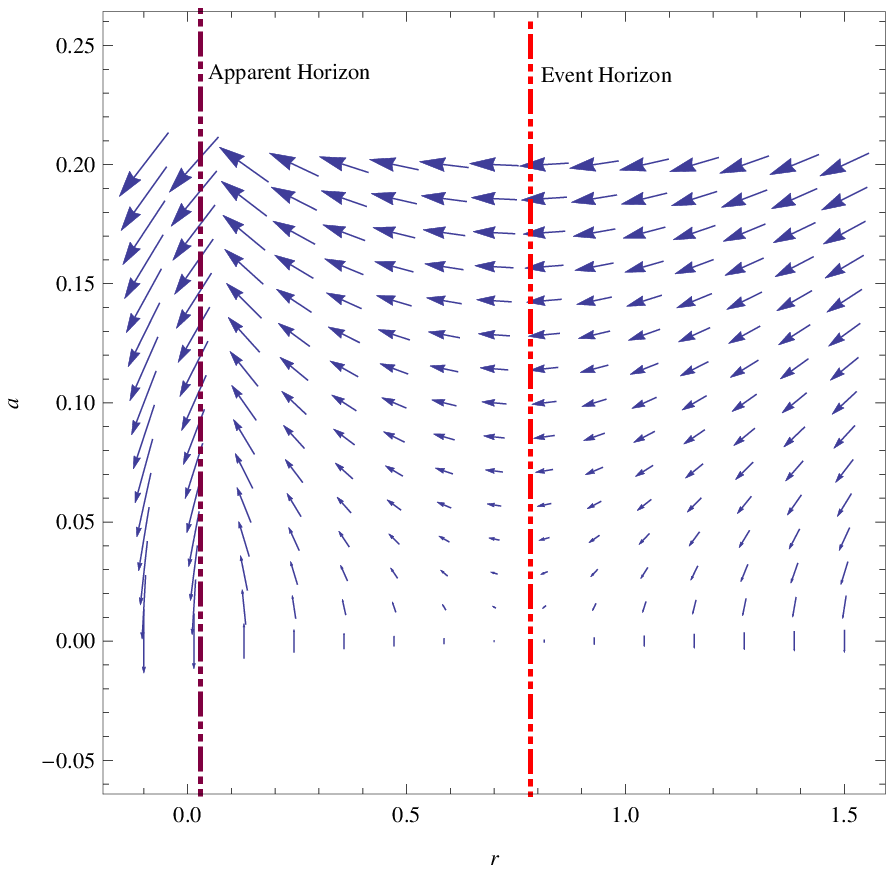}
\caption{\label{fig:7} The near-horizon behavior of the $\bu_2$ vector field.}}
\end{figure}

\begin{figure}[t]
\center{\includegraphics[width=8.5cm]{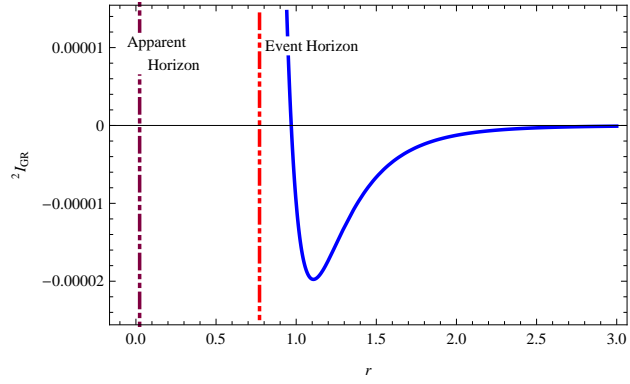}
\caption{\label{fig:8} The congruence intensity $^2I_{\mathrm{GR}}$ for energies given in Fig. \ref{fig:6}.}}
\end{figure}

From Eq. (\ref{eq:XVb}), we have
\begin{equation}\label{eq:XXXII}
u_2^{\alpha }=\left(a,\frac{1}{r}\left(b  \chi^{\frac{1}{2}}+\xi  \chi+M \xi  \chi^{\frac{1}{2}} \ln \left[\chi^{\frac{1}{2}}-M+r\right]\right),0,0\right).
\end{equation}
As before, I calculate the congruence components which now read as
\begin{multline}\label{eq:XXXIII}
^2E_{\mathrm{GR}}=-\frac{Q^2\chi^{-\frac{1}{2}}}{\kappa  r^4 }
\times\left\{\frac{\left(\xi ^2 r^2-a^2\right) \chi^{\frac{3}{2}}}{r^2}+b^2 \chi^{\frac{1}{2}}+2 b \xi  \chi+2 M \xi  \chi^{\frac{1}{2}} \ln \left[\chi^{\frac{1}{2}}-M+r\right] \left(b+\xi  \chi^{\frac{1}{2}}\right)\right.\\
\left.+M^2 \xi ^2 \chi^{\frac{1}{2}} \left(\ln \left[\chi^{\frac{1}{2}}-M+r\right]\right)^2\right\},
\end{multline}
\begin{multline}\label{eq:XXXIV}
^2\Theta _{\text{GR}}=\frac{1}{r^2\chi}
\left[\sqrt{\chi } \left(b \left(-M r+r^2+\chi \right)+\xi  \sqrt{\chi } \left(-M r+2 r^2+\chi \right)\right)\right.\\
\left.+\xi M \sqrt{\chi } \left(-M r+r^2+\chi \right) \ln \left(-M+r+\sqrt{\chi }\right)\right],
\end{multline}
\begin{multline}\label{eq:XXXV}
^2I_{\mathrm{GR}} = \frac{2 Q^2 \chi^{-\frac{1}{2}}}{a \kappa  r^9}
\times \left\{2 b^2 r^2 \chi^{\frac{1}{2}}+b \xi  r^2 \left(r (3 r-8 M)+4 Q^2\right)+\chi^{\frac{1}{2}} \left(a^2 \left(5 M r-3 Q^2-2 r^2\right)\right.\right.\\
\left.\left.+\xi ^2 r^2 \left(-4 M r+2 Q^2+r^2\right)\right)
+M \xi  r^2 \ln \left[\chi^{\frac{1}{2}}-M+r\right] \left(4 b \chi^{\frac{1}{2}}+\xi  \left(-8 M r+4 Q^2+3 r^2\right)\right)\right.\\
\left.+2 M^2 \xi ^2 r^2 \chi^{\frac{1}{2}} \left(\ln \left[\chi^{\frac{1}{2}}-M+r\right]\right)^2\right\}.
\end{multline}
The energy and expansion of the $\bu_2$ congruence have been plotted in Fig. \ref{fig:6}. As it is observed, the dominant energy is the energy due to the charged massive source and is positive. This corresponds to a divergent infalling congruence toward the black hole's event horizon. Moreover, a maximum in the energy at the vicinity of the event horizon, is relevant to a minimum in the expansion which is followed by a complete dispersion on the event horizon. This situation is thoroughly different from what we saw earlier for the congruence of charged particles and also for the case of $\bu_1$ congruence; the $\bu_2$ congruence, because of its dispersion on the horizon, cannot enter the black hole region. However, as it can be seen in Fig.~\ref{fig:7}, the $\bu_2$ vector field is present in the whole region. The figure illustrates the shifts in the $\bu_2$ field behavior around the horizons. The fact that the congruence energy and expansion are not present inside the black hole's event horizon, stems in the elimination of the congruence's kinematical components (as well as the congruence deviation fields). We therefore cannot define the $\bu_2$ "congruence" inside the event horizon, although the $\bu_2$ vector field is well-defined everywhere. The $\bu_2$ congruence has the particular intensity illustrated in Fig.~\ref{fig:8}. The figure shows a switch between negative ({\textit{due to repulsion}}) and positive ({\textit{due to attraction}}) values at the vicinity of the event horizon. Comparing Figs.~\ref{fig:6} and \ref{fig:8}, we can see that the minimum in the divergence, corresponds to a maximum in the repulsive intensity (i.e. smaller negative values). This is because the congruence components are receding from each other at a lower rate. The congruence intensity, then raises remarkably into positive values as the congruence is attracted toward the event horizon, because of the dominance of the gravitational energy in that region.    

\subsubsection{The $u_3$ Congruence}
\begin{figure}[t]
\center{\includegraphics[width=8.3cm]{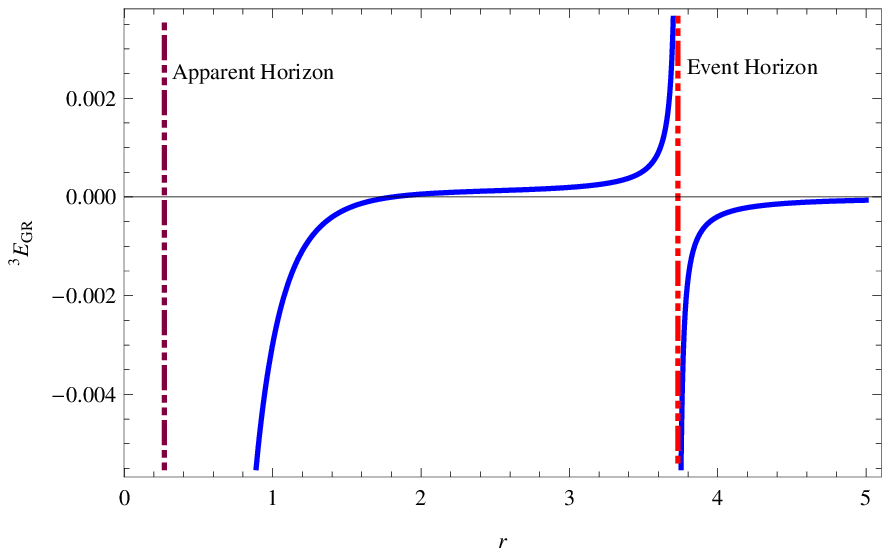}~(a)
\hfil
\includegraphics[width=8.3cm]{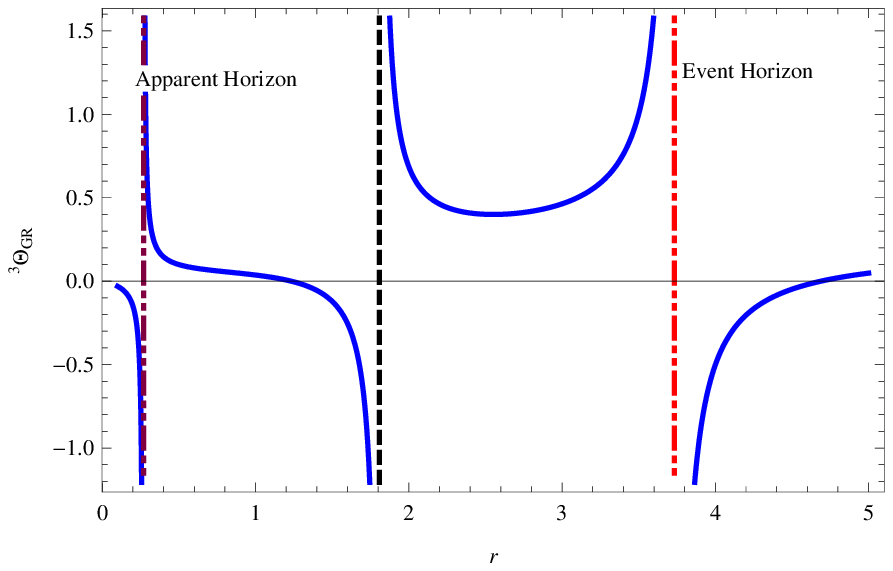}~(b)
\caption{\label{fig:9} The congruence characteristics for $\bu_3$. The figures show (a) the energy and (b) the expansion for the values $a = 0.2$, $M=2$,  $Q=1$ and $\xi=0.1$.}}
\end{figure}

\begin{figure}[t]
\center{\includegraphics[width=8.5cm]{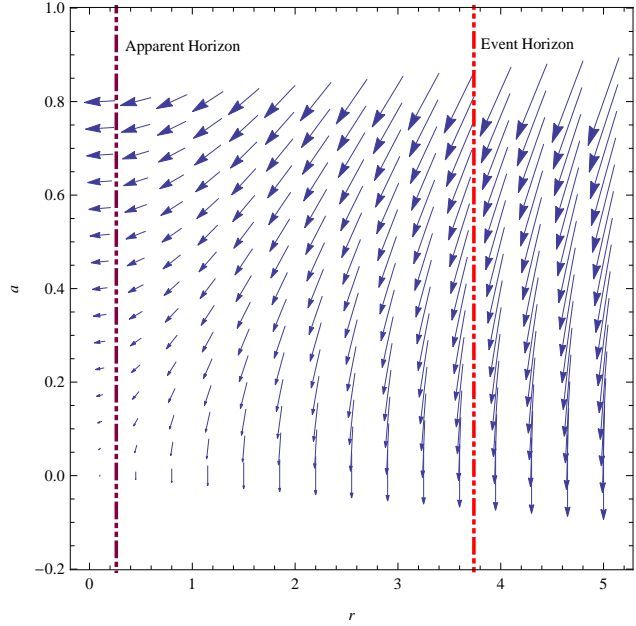}
\caption{\label{fig:10} The near-horizon behavior of the $\bu_3$ vector field.}}
\end{figure}

\begin{figure}[t]
\center{\includegraphics[width=8.5cm]{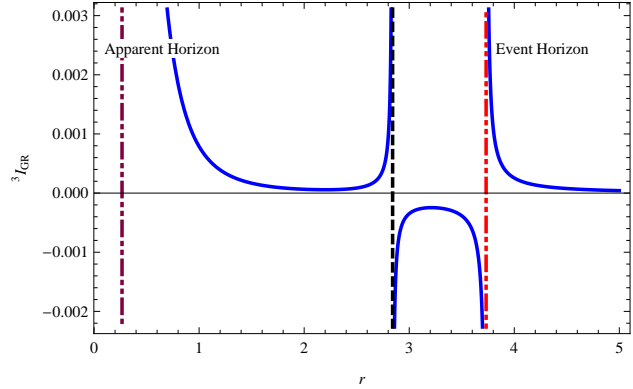}
\caption{\label{fig:11} The congruence intensity $^3I_{\mathrm{GR}}$ for energies given in Fig. \ref{fig:9}.}}
\end{figure}

The $\bu_3$ congruence in Eq. (\ref{eq:XVc}), is indeed independent of the chosen theory of gravity and is therefore, remained unchanged with respect to the alternative solutions for $A(r)$ and $B(r)$. Its kinematical characteristics are calculated as
\begin{equation}\label{eq:XXXVI}
^3E_{\mathrm{GR}}=\frac{Q^2 \left(a^2 \chi ^2-\xi ^2 r^6\right)}{\kappa  r^6 \chi },
\end{equation}
\begin{equation}\label{eq:XXXVII}
^3\Theta_{\mathrm{GR}} = \frac{\xi  r \left(\xi ^2 r^5 \left(r (5 M-2 r)-3 Q^2\right)-a^2 \chi ^2 (M-r)\right)}{\chi  \left(a^2 \chi ^2-\xi ^2 r^6\right)},
\end{equation}
\begin{equation}\label{eq:XXXVIII}
^3I_{\mathrm{GR}} = \frac{2 \xi  Q^2 \left[\chi ^2 a^2 \left(r (2 r-5 M)+3 Q^2\right)+\xi ^2 r^7 (M-r)\right]}{\kappa  r^4 \chi  \left[r^2 \left(4 M^2-4 M r-\xi ^2 r^4+r^2\right)+2 Q^2 r (r-2 M)+Q^4\right]}.
\end{equation}
As illustrated in Figs. \ref{fig:9}, outside the event horizon, the congruence is subjected to gravitational energies, causing it to focus on the horizon. This focusing phenomena, which is followed by a dispersion around the event horizon, corresponds to a singularity in the energy. In other words, the event horizon itself, is a singularity in the congruence energy which makes the matter energy to be dominant inside the black hole region and at the vicinity of the event horizon. This causes a divergent congruence until it reaches the singularity in the expansion. This is the point, where the dominant energy shifts from the matter energy to the gravitational energy. As the energy becomes completely gravitational and of large values, the congruence becomes dispersed on the apparent horizon. This loss of integrity, has been shown in Fig.~\ref{fig:10} in which the $\bu_3$ vector field has been demonstrated. Furthermore, as shown in Fig.~\ref{fig:11}, the congruence is attractive outside the event horizon and at the vicinity of the apparent horizon. This is mostly a correspondent of a convergent congruence. The congruence intensity shows a repulsion for positive energies (where the congruence is divergent), right after the congruence passes the event horizon.

\subsection{Discussion}
The above three different cases, show rather peculiar behaviors for infalling particle congruences on a black hole. Indeed, it is expected that infalling congruences experience convergent conditions outside event horizons, similar to what I discussed for the case of an infalling charged congruence. In the context of general relativity, and since the strong energy condition is satisfied, gravity is always attractive \cite{Poisson2009}. This is what which makes the mentioned convergence, as demonstrated at the beginning of this section for a charged particle congruence in the RN geometry. The homothetic congruences discussed in this section, show however some new behaviors. Since the congruences are uncharged, we do not need to consider the physical acceleration in the common sense. However, the charged sources' effect appears in the spacetime geometry and most importantly, from the gravitational field equations. These concepts, when come together for the above discussed homothetic fields, cause them to be able to behave similar to what are well-known for the charged particles' trajectories around RN black holes. These cast themselves as the main motion categories, known as {\textit{planetary motion}}, {\textit{escaping}} and {\textit{trapping}} \cite{Fathi2013}.\\ 

Let me explain this in details. Regarding the $\bu_1$ congruence expansion in Fig.~\ref{fig:3}, one can observe that in general, the congruence becomes completely attracted to the massive source. This is similar to the trapping fate in the context of particles' motion. For the $\bu_2$ congruence, the whole scenario is different. In Figs.~\ref{fig:6}, we can see that the congruence is dispersed on the horizon and does not enter it\footnote{Note that, the $\bm v$ congruence discussed earlier in this section, does not obey an escaping motion, because it enters the black hole region.}. This is similar to the escaping motion for charged particles' motion. And finally, for the $\bu_3$ congruence, the behavior in Figs.~\ref{fig:9}, shows a frequent approaching-then-receding behavior during the congruence evolution, which is an analog of planetary orbits for charged particles. We therefore can observe that, these congruences' evolution is diverse, and recommend notable alternatives to the usual expectation.

\section{Conclusion}\label{SecV}
 
The evolution, kinematics and physical characteristics of homothetic congruences of uncharged particles, while they fall onto a static charged black hole, constituted the main subject of this paper. In general, I showed that the homotehtic property in the context of a chosen solution to the gravitational field equations, could provide us three different congruence generators, each of which, exhibiting a peculiar evolution within the general relativistic limit and in RN geometry. It turned out that these homothetic fields reveal different results from what is expected; for example, the homothetic congruences can experience a full dispersion outside the black hole's event horizon. 

The whole discussion was based on the kinematical decomposition of the congruences' transverse subspace and a generalized Raychaudhuri equation. The task was elaborated by making benefit from the imposition of the gravitational correspondent of the source's energy-momentum tensor into the kinematical components. This way, one can examine the effects of the non-gravitational energies on the congruences in a given theory of gravity, by only making use of the spacetime solutions.

In the current study, these non-gravitational effects come from an electrically charged massive source of gravity and the generalized kinematics, enabled the examination of the effects of this charged source on the uncharged congruences. The three cases of congruences, were indeed different in physics, evolution and fate. Whereas, classically, one expects a steady convergence for infalling congruences onto a black hole's horizon, the uncharged homothetic congruences could encounter a primary dispersion (infinite positive expansion) and/or focusing (infinite negative expansion), outside and/or inside the black hole's region. As discussed in the previous section, these behaviors, when combined together in certain orders, constitute conditions similar to the three main categories of particles' motion in general relativity. I therefore conclude that the homothehic property and the association of congruence kinematics with the energy-momentum deposit of the gravitational source, enables us to tie the concepts of {\textit{particle trajectories}} and {\textit{congruence evolution}} in general relativistic investigations. According to the fact that the particles' trajectories are of great interests in observational tests of gravitational theories, the generalized kinematics and the methods introduced here, can be considered as useful tools for making theoretical predictions for those tests in alternative theories of gravity.

\appendix

\section{The Congruence Deviation Relation for Homothetic Fields}\label{app:appendixA}

The evolution of the $\bJ\cdot\bu$ scalar is calculated as
\begin{eqnarray}\label{eq:appAI}
(\dot{\bJ\cdot\bu}) &=& (J^\alpha u_\alpha)_{;\beta}u^\beta\nonumber\\
&=& J^\alpha{_{;\beta}}u^\beta u_\alpha + J^\alpha u_{\alpha;\beta} u^\beta = J^\alpha{_{;\beta}}u^\beta u_\alpha + J^\alpha a_\alpha\nonumber\\
&=&  J^\alpha{_{;\beta}}u^\beta u_\alpha - \frac{1}{2}J^\alpha (\bu\cdot\bu)_{;\alpha} + 2\xi J^\alpha u_\alpha\nonumber\\
&=& u^\alpha{_{;\beta}}J^\beta u_\alpha - \frac{1}{2}J^\alpha (\bu\cdot\bu)_{;\alpha} + 2\xi J^\alpha u_\alpha\nonumber\\
&=& \frac{1}{2}(\bu\cdot\bu)_{;\beta}J^\beta - \frac{1}{2}J^\alpha(\bu\cdot\bu)_{;\alpha}+2\xi J^\alpha u_\alpha\nonumber\\
&=& 2 \xi (\bJ\cdot\bu),
\end{eqnarray}
in which I have used the condition in Eq. (\ref{eq:IV}) in the fourth line and in the fifth line, the homothetic acceleration in Eq. (\ref{eq:III}) has been interpolated. The second order evolution of the Jacobi fields, will result in the congruence deviation equation. We have
\begin{eqnarray}\label{eq:appAII}
\ddot J^\alpha &=& (J^\alpha{_{;\beta}}u^\beta)_{;\gamma}u^\gamma 
= (u^\alpha{_{;\beta}}J^\beta)_{;\gamma}u^\gamma = u^\alpha{_{;\beta\gamma}}J^\beta u^\gamma + u^\alpha{_{;\beta}}J^\beta{_{;\gamma}}u^\gamma\nonumber\\
&=& (u^\alpha{_{;\gamma\beta}}-R^\alpha{_{\mu\beta\gamma}}u^\mu)
J^\beta u^\gamma + u^\alpha{_{;\beta}}u^\beta{_{;\gamma}}J^\gamma\nonumber\\
&=& (u^\alpha{_{;\gamma}}u^\gamma)_{;\beta}J^\beta - u^\alpha{_{;\gamma}}u^\gamma{_{;\beta}}J^\beta - R^\alpha{_{\mu\beta\gamma}}u^\mu J^\beta u^\gamma + u^\alpha{_{;\beta}}u^\beta{_{;\gamma}}J^\gamma\nonumber\\
&=& a^\alpha{_{;\beta}}J^\beta - R^\alpha{_{\mu\beta\gamma}}u^\mu J^\beta u^\gamma\nonumber\\
&=& \left(-\frac{1}{2}(\bu\cdot\bu)^{;\alpha}{_{;\beta}}+2 \xi u^\alpha{_{;\beta}}\right) J^\beta - R^\alpha{_{\mu\beta\gamma}}u^\mu J^\beta u^\gamma\nonumber\\
&=& -\frac{1}{2}g^{\alpha\lambda}\left[\nabla_\beta\nabla_\lambda\left(\bu\cdot\bu\right)\right]J^\beta
+2 \xi u^\alpha{_{;\beta}}J^\beta - R^\alpha{_{\mu\beta\gamma}}u^\mu J^\beta u^\gamma\nonumber\\
&=& -\frac{1}{2}g^{\alpha \lambda } (\bu\cdot\bu)_{,\lambda \beta } J^\beta + 2 \xi  \dot{J}^{\alpha }-R^{\alpha }{}_{\mu \beta \gamma } u^{\mu } J^{\beta } u^{\gamma }.
\end{eqnarray}

\section{Derivation of the Generalized Raychaudhuri Equation}\label{appII}

The derivation carried out here, is the time-like counterpart of that, which has been given in Ref. \cite{Thompson2017}. Expanding Eqs. (\ref{eq:dotbarB}) and (\ref{eq:barB-expansion}), we have 
\begin{eqnarray}\label{eq:dotTheta-1}
\dot\Theta &=& \frac{D}{\ed\tau}\bar{B}{^\mu}_\mu = \frac{D}{\ed\tau}\left(h{_\rho}^\mu u{^\rho}_{;\beta} h{_\mu}^\beta\right)\nonumber\\
&=&h{_\rho}^\mu\frac{D}{\ed\tau}\left(u{^\rho}_{;\beta}\right)h{_\mu}^\beta + \frac{D}{\ed\tau}\left(h{_\rho}^\mu h{_\mu}^\beta\right)u{^\rho}_{;\beta}\nonumber\\
&=&h{_\rho}^\mu\left(u^\tau u{^\rho}_{;\tau\beta}-R{^\rho}_{\lambda\beta\tau}u^\lambda u^\tau\right)h{_\mu}^\beta+u^\tau\left(
h{_\rho}^\mu h{_\mu}^\beta
\right)_{;\tau} u{^\rho}_{;\beta}\nonumber\\
&=& h{_\rho}^\mu \left(
a{^\rho}_{;\beta}-u{^\rho}_{;\tau} u{^\tau}_{;\beta}
-R{^\rho}_{\lambda\beta\tau} u^\lambda u^\tau
\right)h{_\mu}^\beta + u^\tau\left(
h{_\rho}^\beta
\right)_{;\tau}u{^\rho}_{;\beta}\nonumber\\
&=& h{_\rho}^\beta a{^\rho}_{;\beta}+h{_\rho}^\mu u{^\rho}_{;\tau}u{^\tau}_{;\beta}h{_\mu}^\beta - h{_\rho}^\beta R{^\rho}_{\lambda\beta\tau}u^\lambda u^\tau + \dot{h}{_\rho}^\beta u{^\rho}_{;\beta},
\end{eqnarray}
in which 
\begin{eqnarray}\label{eq:firstTerm}
\dot{h}{_\rho}^\beta &=& -\frac{D}{\ed\tau}\left[(\bm p\cdot\bm u)^{-1}\right]p_\alpha u^\beta - (\bm p\cdot \bm u)^{-1}\left(\dot{p}_\rho u^\beta + p_\rho a^\beta\right)\nonumber\\
&=&(\bm p\cdot \bm u)^{-2}\dot{(\bm p\cdot \bm u)}p_\rho u^\beta -(\bm p\cdot\bm u)^{-1}\left(\dot{p}_\rho u^\beta + p_\rho a^\beta\right).
\end{eqnarray}
Now let me expand the terms individually. 
\begin{eqnarray}\label{eq:1stTerm}
{h}{_\rho}^\beta a{^\rho}_{;\beta} &=& a{^\rho}_{;\rho} - (\bm p\cdot \bm u)^{-1}\left(p_\rho u^\beta\right)a{^\rho}_{;\beta}\nonumber\\
&=&a{^\rho}_{;\rho}-(\bm p\cdot \bm u)^{-1}(\bm p \cdot \dot{\bm a}).
\end{eqnarray}
\begin{eqnarray}\label{eq:2ndTerm}
h{_\rho}^\mu u{^\rho}_{;\tau} u{^\tau}_{;\beta} h{_\mu}^\beta &=&
\left(h{_\rho}^\mu u{^\rho}_{;\tau}h{_\nu}^\tau\right)\left(
h{_\sigma}^\nu u{^\sigma}_{;\beta} h{_\mu}^\beta\right) +
h{_\rho}^\mu u{^\rho}_{;\tau}\left(\delta^\tau_\nu - h{_\nu}^\tau\right)
\left(\delta_\sigma^\nu-h{_\sigma}^\nu\right)u{^\sigma}_{;\beta} h{_\mu}^\beta\nonumber\\
&=& \bar{B}{^\mu}_\nu\bar{B}{^\nu}_\mu + h{_\rho}^\mu u{^\rho}_{;\tau}
\left(\delta^\tau_\nu - h{_\nu}^\tau\right)
\left(\delta_\sigma^\nu-h{_\sigma}^\nu\right)u{^\sigma}_{;\beta} h{_\mu}^\beta.
\end{eqnarray}
\begin{eqnarray}\label{eq:3rdTerm}
h{_\rho}^\beta R{^\rho}_{\lambda\beta\tau} &=& R_{\lambda\tau}u^\lambda u^\tau - (\bm p\cdot \bm u)^{-1}\left(\underbrace{p_\rho u^\beta R{^\rho}_{\lambda\beta\tau}}_{=0}\right) = R_{\lambda\tau} u^\lambda u^\tau.
\end{eqnarray}
Applying the expressions in Eqs. (\ref{eq:1stTerm}), (\ref{eq:2ndTerm}) and (\ref{eq:3rdTerm}) in Eq. (\ref{eq:dotTheta-1}), we get
\begin{eqnarray}\label{eq:1stEq}
\dot\Theta &=& -\bar{B}{^\mu}_\nu \bar{B}{^\nu}_\mu - R_{\lambda\tau}u^\lambda u^\tau + a{^\rho}_{;\rho} - (\bm p\cdot \bm u)^{-1}(\bm p\cdot \dot{\bm a}) \nonumber\\
&&- h{_\rho}^\mu u{^\rho}_{;\tau}\left(\delta^\tau_\nu - h{_\nu}^\tau\right)
\left(\delta_\sigma^\nu-h{_\sigma}^\nu\right)u{^\sigma}_{;\beta} h{_\mu}^\beta + \dot h{_\rho}^\beta u{^\rho}_{;\beta}\nonumber\\\nonumber\\
&=&-\bar{B}{^\mu}_\nu \bar{B}{^\nu}_\mu - R_{\lambda\tau}u^\lambda u^\tau + a{^\rho}_{;\rho} - (\bm p\cdot \bm u)^{-1}(\bm p\cdot \dot{\bm a}) \nonumber\\
&&- h{_\rho}^\beta u{^\rho}_{;\tau}\left(\delta^\tau_\sigma - h{_\sigma}^\tau\right)u{^\sigma}_{;\beta} + \dot h{_\rho}^\beta u{^\rho}_{;\beta}.
\end{eqnarray}
Now consider the term $h{_\rho}^\beta u{^\rho}_{;\tau}\left(\delta_\sigma^\tau - h{_\sigma}^\tau\right)u{^\sigma}_{;\beta}\equiv h{_\rho}^\mu u{^\rho}_{;\tau}u{^\nu}_{;\mu}\left(\delta_\nu^\tau - h{_\nu}^\tau\right)$, which gives
\begin{eqnarray}\label{eq:2ndEq}
h{_\rho}^\mu \left(\delta_\nu^\tau - h{_\nu}^\tau\right) u{^\nu}_{;\mu} u{^\rho}_{;\tau} &=& \left[
\delta_\rho^\mu - (\bm p\cdot \bm u)^{-1}p_\rho u^\mu
\right]
\left[
(\bm p\cdot \bm u)^{-1} p_\nu u^\tau
\right]u{^\nu}_{;\mu}u{^\rho}_{;\tau}\nonumber\\
&=& (\bm p\cdot \bm u)^{-1} p_\nu u^\tau \delta_\rho^\mu u{^\nu}_{;\mu} u{^\rho}_{;\tau} - (\bm p\cdot \bm u)^{-2}p_\rho p_\nu u^\mu u^\tau u{^\nu}_{;\mu} u{^\rho}_{;\tau}\nonumber\\
&=& (\bm p\cdot \bm u)^{-1} p_\nu u^\tau u{^\nu}_{\rho}u{^\rho}_{;\tau} - (\bm p\cdot \bm u)^{-2} p_\rho p_\nu a^\nu a^\rho\nonumber\\
&=& (\bm p\cdot \bm u)^{-1} p_\nu u{^\nu}_{;\rho}a^\rho - (\bm p\cdot \bm u)^{-2} (\bm p\cdot \bm a)^2.
\end{eqnarray}
On the other hand,
\begin{eqnarray}\label{eq:3rdEq}
\dot{h}{_\rho}^\beta u{^\rho}_{;\beta} &=& \left[
(\bm p\cdot \bm u)^{-2}\dot{(\bm p\cdot \bm u)}p_\rho u^\beta - (\bm p\cdot \bm u)^{-1} \left(
\dot p_\rho u^\beta + p_\rho a^\beta\right)
\right] u{^\rho}_{;\beta}\nonumber\\
&=& (\bm p\cdot \bm u)^{-2} \dot{(\bm p\cdot \bm u)} (\bm p\cdot \bm a) 
- (\bm p\cdot \bm u)^{-1}\left(\dot{\bm p}\cdot \bm a + p_\rho u{^\rho}_{;\beta} a^\beta\right).
\end{eqnarray}
Interpolating Eqs. (\ref{eq:2ndEq}) and (\ref{eq:3rdEq}) into Eq. (\ref{eq:1stEq}), we obtain
\begin{eqnarray}\label{eq:Raycahudhuri-1}
\dot{\Theta} &=& -\bar{B}{^\mu}_\nu\bar{B}{^\nu}_\mu - R_{\lambda\tau}u^\lambda u^\tau + a{^\rho}_{;\rho} - (\bm p\cdot \bm u)^{-1} (\bm p\cdot \dot{\bm a})\nonumber\\
&&- (\bm p \cdot \bm u)^{-1}p_\nu u{^\nu}_{;\rho}a^\rho + (\bm p\cdot \bm u)^{-2} (\bm p\cdot \bm a)^2\nonumber\\
&&+(\bm p \cdot \bm u)^{-2} \dot{(\bm p\cdot \bm u)}(\bm p\cdot \bm a)
-(\bm p\cdot \bm u)^{-1}\left((\dot{\bm p}\cdot \bm a) + p_\rho u{^\rho}_{;\beta} a^\beta\right)\nonumber\\\nonumber\\
&=& -\bar{B}{^\mu}_\nu\bar{B}{^\nu}_\mu - R_{\lambda\tau}u^\lambda u^\tau + a{^\rho}_{;\rho}\nonumber\\
&&- (\bm p\cdot \bm u)^{-1}\left[
\dot{(\bm p\cdot \bm a)} + 2 p_\nu u{^\nu}_{;\rho} a^\rho
\right]\nonumber\\
&&+(\bm p\cdot\bm u)^{-2}\left[
(\bm p\cdot \bm a)^2+\dot{(\bm p\cdot \bm u)}(\bm p\cdot \bm a)
\right].
\end{eqnarray}
I introduce the decomposition of the transverse tensor $\bar{B}{^\mu}_\nu$ into its symmetric and anti-symmetric parts as
\begin{equation}\label{eq:barB-decomposed}
\bar{B}{^\mu}_\nu = \theta{^\mu}_\nu + \omega{^\mu}_\nu,
\end{equation}
in which the symmetric tensor $\theta{^\mu}_\nu$ itself, can be decomposed into trace and trace-less parts as 
\begin{equation}\label{eq:theta-decomposed}
\theta{^\mu}_\nu = \frac{1}{3}\Theta h{_\nu}^\mu + \sigma{^\mu}_\nu.
\end{equation}
Having these, one can simply justify that
\begin{equation}\label{eq:B.B-decomposed}
\bar{B}{^\mu}_\nu\bar{B}{^\nu}_\mu = \frac{1}{3}\Theta^2 + \sigma{^\mu}_\nu\sigma{^\nu}_\mu + \omega{^\mu}_\nu\omega{^\nu}_\mu = \frac{1}{3}\Theta^2 + \sigma_{\mu\nu}\sigma^{\mu\nu} - \omega_{\mu\nu}\omega^{\mu\nu}.
\end{equation}
Inserting this into Eq. (\ref{eq:Raycahudhuri-1}), we finally reach the generalized Raychaudhuri equation which reads as
\begin{eqnarray}\label{eq:Raycahudhuri-2}
\dot{\Theta} &=&  -\frac{1}{3}\Theta^2 - \sigma_{\mu\nu}\sigma^{\mu\nu} + \omega_{\mu\nu}\omega^{\mu\nu} - R_{\lambda\tau}u^\lambda u^\tau + a{^\rho}_{;\rho}- (\bm p\cdot \bm u)^{-1}\left[
\dot{(\bm p\cdot \bm a)} + 2 p_\nu u{^\nu}_{;\rho} a^\rho
\right]\nonumber\\
&&+(\bm p\cdot\bm u)^{-2}\left[
(\bm p\cdot \bm a)^2+\dot{(\bm p\cdot \bm u)}(\bm p\cdot \bm a)
\right].
\end{eqnarray}
The kinematical quantities, namely the expansion scalar, the shear and vorticity tensors, can be calculated as follows. For the expansion, we have
\begin{eqnarray}\label{eq:Expansion-derivation}
\Theta &=& \bar{B}{^\mu}_\mu = h{_\rho}^\mu u{^\rho}_{;\beta}h{_\mu}^\beta = h{_\rho}^\beta u{^\rho}_{;\beta}\nonumber\\
&=& \left(\delta_\rho^\beta - (\bm p\cdot \bm u)^{-1}p_\rho u^\beta\right)u{^\rho}_{;\beta}\nonumber\\
&=& u{^\beta}_{;\beta} - (\bm p\cdot \bm u)^{-1} (\bm p\cdot \bm a).
\end{eqnarray}
To deal with shear and vorticity, we need to bear in mind that $\theta{^\mu}_\nu$ in Eq. (\ref{eq:theta-decomposed}) is indeed
\begin{equation}\label{eq:Shear-vorticity-derivation-1}
\theta{^\mu}_\nu = \bar{B}_{(\mu\sigma)}g^{\sigma\nu}.
\end{equation}
We therefore may write these two quantities as 
\begin{subequations}\label{eq:Shear-vorticity-derivation-2}
\begin{align}
\sigma_{\mu\nu}=h{_\mu}^\alpha u_{(\alpha;\beta)}h{_\nu}^\beta - \frac{1}{3}\Theta h_{\mu\nu},\label{eq:Shear-vorticity-derivation-2-a}\\
\omega_{\mu\nu}=h{_\mu}^\alpha u_{[\alpha;\beta]}h{_\nu}^\beta,\label{eq:Shear-vorticity-derivation-2-b}
\end{align}
\end{subequations}
where $h_{\mu\nu} = h{_\mu}^\alpha g_{\alpha\nu}$. The special case of  $\bp = m \bu$ with $m = \mathrm{const.}$, offers $\bu\cdot \bu = \mathrm{const.}$ (normalized congruence). The Raychaudhuri equation for such systems is given by reducing the generalized equation in Eq. (\ref{eq:Raycahudhuri-2}). For this case we have
\begin{eqnarray}\label{eq:Raycahudhuri-reduced-1}
\dot{\Theta} &=&  -\frac{1}{3}\Theta^2 - \sigma_{\mu\nu}\sigma^{\mu\nu} + \omega_{\mu\nu}\omega^{\mu\nu} - R_{\lambda\tau}u^\lambda u^\tau + a{^\rho}_{;\rho}\nonumber\\
&&- (\bm u\cdot \bm u)^{-1}\left[
\dot{(\bm u\cdot \bm a)} + 2 u_\nu u{^\nu}_{;\rho} a^\rho
\right]\nonumber\\
&&+(\bm u\cdot\bm u)^{-2}\left[
(\bm u\cdot \bm a)^2+\underbrace{\dot{(\bm u\cdot \bm u)}}_{=0}(\bm u\cdot \bm a)
\right].
\end{eqnarray}
The term $\bm u\cdot \bm a$ gives
\begin{equation}\label{eq:1stTerm-reduction}
\bm u\cdot \bm a = u_\alpha u{^\alpha}_{;\beta} u^\beta = \frac{1}{2}\left(\bm u\cdot\bm u\right)_{;\beta} u^\beta = 0,
\end{equation}
which shows that the terms $u_{\nu} u{^\nu}_{;\rho}a^\rho$ vanishes identically. We therefore come up with the reduced Raychaudhuri equation for normalized congrunces which has the very well-known form
\begin{eqnarray}\label{eq:Raycahudhuri-reduced-2}
\dot{\Theta} &=&  -\frac{1}{3}\Theta^2 - \sigma_{\mu\nu}\sigma^{\mu\nu} + \omega_{\mu\nu}\omega^{\mu\nu} - R_{\lambda\tau}u^\lambda u^\tau + a{^\rho}_{;\rho}.
\end{eqnarray}
By means of Eq.~(\ref{eq:Expansion-derivation}), we get $\Theta = u{^\beta}_{;\beta}$.


\begin{thebibliography}{45}%
\makeatletter
\providecommand \@ifxundefined [1]{%
 \@ifx{#1\undefined}
}%
\providecommand \@ifnum [1]{%
 \ifnum #1\expandafter \@firstoftwo
 \else \expandafter \@secondoftwo
 \fi
}%
\providecommand \@ifx [1]{%
 \ifx #1\expandafter \@firstoftwo
 \else \expandafter \@secondoftwo
 \fi
}%
\providecommand \natexlab [1]{#1}%
\providecommand \enquote  [1]{``#1''}%
\providecommand \bibnamefont  [1]{#1}%
\providecommand \bibfnamefont [1]{#1}%
\providecommand \citenamefont [1]{#1}%
\providecommand \href@noop [0]{\@secondoftwo}%
\providecommand \href [0]{\begingroup \@sanitize@url \@href}%
\providecommand \@href[1]{\@@startlink{#1}\@@href}%
\providecommand \@@href[1]{\endgroup#1\@@endlink}%
\providecommand \@sanitize@url [0]{\catcode `\\12\catcode `\$12\catcode
  `\&12\catcode `\#12\catcode `\^12\catcode `\_12\catcode `\%12\relax}%
\providecommand \@@startlink[1]{}%
\providecommand \@@endlink[0]{}%
\providecommand \url  [0]{\begingroup\@sanitize@url \@url }%
\providecommand \@url [1]{\endgroup\@href {#1}{\urlprefix }}%
\providecommand \urlprefix  [0]{URL }%
\providecommand \Eprint [0]{\href }%
\providecommand \doibase [0]{http://dx.doi.org/}%
\providecommand \selectlanguage [0]{\@gobble}%
\providecommand \bibinfo  [0]{\@secondoftwo}%
\providecommand \bibfield  [0]{\@secondoftwo}%
\providecommand \translation [1]{[#1]}%
\providecommand \BibitemOpen [0]{}%
\providecommand \bibitemStop [0]{}%
\providecommand \bibitemNoStop [0]{.\EOS\space}%
\providecommand \EOS [0]{\spacefactor3000\relax}%
\providecommand \BibitemShut  [1]{\csname bibitem#1\endcsname}%
\let\auto@bib@innerbib\@empty




\bibitem [{\citenamefont {Misner}\ \emph {et~al.}(1973)\citenamefont {Misner},
  \citenamefont {Thorne},\ and\ \citenamefont {Wheeler}}]{MTW1973}%
  \BibitemOpen
  \bibfield  {author} {\bibinfo {author} {\bibfnamefont {C.~W.}\ \bibnamefont
  {Misner}}, \bibinfo {author} {\bibfnamefont {K.~S.}\ \bibnamefont {Thorne}},
  \ and\ \bibinfo {author} {\bibfnamefont {J.~A.}\ \bibnamefont {Wheeler}},\
  }\href@noop {} {\emph {\bibinfo {title} {Gravitation}}}\ (\bibinfo
  {publisher} {Freeman},\ \bibinfo {address} {San Francisco},\
  \bibinfo {year} {1973})\BibitemShut {NoStop}%
  \bibitem [{\citenamefont {Gauss}(1986)}]{Gauss1986}%
  \BibitemOpen
  \bibfield  {author} {\bibinfo {author} {\bibfnamefont {C. F.}\ \bibnamefont
  {Gauss}} and \bibinfo {author} {\bibfnamefont {W. C.}\ \bibnamefont
  {Waterhouse}},
       {\href {\doibase 10.1007/978-1-4939-7560-0} {\bibfield  {chapter}
  {\bibinfo  {chapter} \emph{Disquisitiones Arithmeticae}}}}
  {\  (\bibinfo  {publisher}{Springer-Verlag New York},\
  \bibinfo {year} {1986})}
 } \BibitemShut {NoStop}%
  \bibitem [{\citenamefont {Hawking}(1973)}]{Hawking1973}%
  \BibitemOpen
  \bibfield  {author} {\bibinfo {author} {\bibfnamefont {S. W.}\ \bibnamefont
  {Hawking}} and \bibinfo {author} {\bibfnamefont {G. F. R.}\ \bibnamefont
  {Ellis}},
       {  {\bibfield  {chapter}
  {\bibinfo  {chapter} \emph{The Large Scale Structure of Space-Time}}}}
  {\  (\bibinfo  {publisher}{Cambridge University Press},\
  \bibinfo {year} {1973})}
 } \BibitemShut {NoStop}%
   \bibitem [{\citenamefont {Penrose}(1965)}]{Penrose1965}%
  \BibitemOpen
  \bibfield  {author} {\bibinfo {author} {\bibfnamefont {R.}\ \bibnamefont
  {Penrose}},
       \href {\doibase 10.1103/PhysRevLett.14.57} {\bibfield  {journal}
  {\bibinfo  {journal} {Phys. Rev. Lett.}\ }\textbf {\bibinfo {volume} {14}},\ \bibinfo{issueNo.}{57}  (\bibinfo {year} {1965})}}\BibitemShut {NoStop}%
  \bibitem [{\citenamefont {Hawking}(1965)}]{Hawking1965}%
  \BibitemOpen
  \bibfield  {author} {\bibinfo {author} {\bibfnamefont {S. W.}\ \bibnamefont
  {Hawking}},
       \href {\doibase 10.1103/PhysRevLett.15.689} {\bibfield  {journal}
  {\bibinfo  {journal} {Phys. Rev. Lett.}\ }\textbf {\bibinfo {volume} {15}},\ \bibinfo{issueNo.}{689}  (\bibinfo {year} {1965})}}\BibitemShut {NoStop}%
  \bibitem [{\citenamefont {Hawking}(1966)}]{Hawking1966}%
  \BibitemOpen
  \bibfield  {author} {\bibinfo {author} {\bibfnamefont {S. W.}\ \bibnamefont
  {Hawking}},
       \href {\doibase 10.1103/PhysRevLett.17.444} {\bibfield  {journal}
  {\bibinfo  {journal} {Phys. Rev. Lett.}\ }\textbf {\bibinfo {volume} {17}},\ \bibinfo{issueNo.}{444}  (\bibinfo {year} {1966})}}\BibitemShut {NoStop}%
 \bibitem [{\citenamefont {Penrose}(2002)}]{Penrose2002}%
  \BibitemOpen
  \bibfield  {author} {\bibinfo {author} {\bibfnamefont {R.}\ \bibnamefont
  {Penrose}},
       \href {\doibase 0001-7701/02/0700-1141/0} {\bibfield  {journal}
  {\bibinfo  {journal} {Gen. Rel. Gravit.}\ }\textbf {\bibinfo {volume} {34}},\ \bibinfo{issueNo.}{7}  (\bibinfo {year} {2002})}}\BibitemShut {NoStop}%
  \bibitem [{\citenamefont {Raychaudhuri}(1955)}]{Raychaudhuri1955}%
  \BibitemOpen
  \bibfield  {author} {\bibinfo {author} {\bibfnamefont {A.}\ \bibnamefont
  {Raychaudhuri}},
       \href {\doibase 10.1103/PhysRev.98.1123} {\bibfield  {journal}
  {\bibinfo  {journal} {Phys. Rev.}\ }\textbf {\bibinfo {volume} {98}},\ \bibinfo{issueNo.}{1123}  (\bibinfo {year} {1955})}}\BibitemShut {NoStop}
  \bibitem [{\citenamefont {Poisson}(2009)}]{Poisson2009}%
  \BibitemOpen
  \bibfield  {author} {\bibinfo {author} {\bibfnamefont {E.}\ \bibnamefont
  {Poisson}},
       {\href {\doibase 10.1017/CBO9780511606601} {\bibfield  {chapter}
  {\bibinfo  {chapter} \emph{A Relativist's Toolkit: The Mathematics of Black-Hole Mechanics}}}}
  {
  \  (\bibinfo  {publisher} {Cambridge University Press},\
  \bibinfo {year} {2009})}
 } \BibitemShut {NoStop}%
  \bibitem [{\citenamefont {Ellis}(2007)}]{Ellis2007}%
  \BibitemOpen
  \bibfield  {author} {\bibinfo {author} {\bibfnamefont {G. F. R.}\ \bibnamefont
  {Ellis}},
       \href {\doibase 10.1007/s12043-007-0107-4} {\bibfield  {journal}
  {\bibinfo  {journal} {Pramana}\ }\textbf {\bibinfo {volume} {69}},\ \bibinfo{issueNo.}{1}, \bibinfo{pages}{15}   (\bibinfo {year} {2007})}}\BibitemShut {NoStop}
   \bibitem [{\citenamefont {Kar}(2007)}]{Kar2007}%
  \BibitemOpen
  \bibfield  {author} {\bibinfo {author} {\bibfnamefont {S.}\ \bibnamefont
  {Kar}},
  \bibinfo {author} {\bibfnamefont {S.}\ \bibnamefont
  {Sengupta}},
       \href {\doibase 10.1007/s12043-007-0110-9} {\bibfield  {journal}
  {\bibinfo  {journal} {Pramana}\ }\textbf {\bibinfo {volume} {69}},\ \bibinfo{issueNo.}{1}, \bibinfo{pages}{49}   (\bibinfo {year} {2007})}}\BibitemShut {NoStop}
 \bibitem [{\citenamefont {Stephani}\ \emph {et~al.}(2003)\citenamefont {Stephani},
  \citenamefont {Kramer}, \citenamefont {MacCallum},  \citenamefont {Hoenselaers}\ and\ \citenamefont {Herlt}}]{Stephani2003}%
  \BibitemOpen
  \bibfield  {author} {\bibinfo {author} {\bibfnamefont {H.}\ \bibnamefont
  {Stephani}}, \bibinfo {author} {\bibfnamefont {D.}\ \bibnamefont {Kramer}},
  \bibinfo {author} {\bibfnamefont {M. A. H.}\ \bibnamefont {MacCallum}},
  \bibinfo {author} {\bibfnamefont {C.}\ \bibnamefont {Hoenselaers}},
  \ and\ \bibinfo {author} {\bibfnamefont {E.}\ \bibnamefont {Herlt}},\
  }\href@noop {} {\emph {\bibinfo {title} {Exact Solutions of Einstein's
Field Equations}}}\ (\bibinfo
  {publisher} {Cambridge University Press},\ 
  \bibinfo {year} {2003})\BibitemShut {NoStop}%
   \bibitem [{\citenamefont {Penrose}(1999)}]{Penrose1999}%
  \BibitemOpen
  \bibfield  {author} {\bibinfo {author} {\bibfnamefont {R.}\ \bibnamefont
  {Penrose}},
       \href {\doibase 10.1007/BF02702355} {\bibfield  {journal}
  {\bibinfo  {journal} {J. Astrophys. Astron.}\ }\textbf {\bibinfo {volume} {20}},\ \bibinfo{issueNo.}{3-4}, \bibinfo{pages}{233}  (\bibinfo {year} {1999})}}\BibitemShut {NoStop}
  \bibitem [{\citenamefont {Faraoni}(2015)}]{Faraoni2015}%
  \BibitemOpen
  \bibfield  {author} {\bibinfo {author} {\bibfnamefont {V.}\ \bibnamefont
  {Faraoni}},
       {\href {\doibase 10.1007/978-3-319-19240-6
} {\bibfield  {chapter}
  {\bibinfo  {chapter} \emph{Cosmological and Black Hole Apparent Horizons}}}}
  {
  \  (\bibinfo{publisher}{Springer},\
  \bibinfo {year} {2015})}
 } \BibitemShut {NoStop}%
    \bibitem [{\citenamefont {Hall}(1988)}]{Hall1988}%
  \BibitemOpen
  \bibfield  {author} {\bibinfo {author} {\bibfnamefont {G. S.}\ \bibnamefont
  {Hall}},
       \href {\doibase 10.1088/0264-9381/5/5/001} {\bibfield  {journal}
  {\bibinfo  {journal} {Class. Quantum Grav.}\ }\textbf {\bibinfo {volume} {5}},\ \bibinfo{issueNo.}{5}, \bibinfo{pages}{L77}  (\bibinfo {year} {1988})}}\BibitemShut {NoStop}
  \bibitem [{\citenamefont {Hall}(2004)}]{Hall2004}%
  \BibitemOpen
  \bibfield  {author} {\bibinfo {author} {\bibfnamefont {G. S.}\ \bibnamefont
  {Hall}},
       {{}{\bibfield  {chapter}
  {\bibinfo  {chapter} \emph{Symmetries and Curvature Structure in General Relativity
}}}}{\  (\bibinfo{publisher}{World Scientific},\
  \bibinfo {year} {2004})}
 } \BibitemShut {NoStop}%
     \bibitem [{\citenamefont {Ahmad}(2017)}]{Ahmad2017}%
  \BibitemOpen
  \bibfield  {author} {\bibinfo {author} {\bibfnamefont {D.}\ \bibnamefont
  {Ahmad}},
  \bibinfo {author} {\bibfnamefont {K.}\ \bibnamefont
  {Habib}},
       \href {\doibase 10.1155/2018/8195208} {\bibfield  {journal}
  {\bibinfo  {journal} {Advances in Mathematical Physics}\ }\textbf {\bibinfo {volume} {2018}},\ \bibinfo{issueNo.}{8195208} (\bibinfo {year} {2018})}}\BibitemShut {NoStop}
      \bibitem [{\citenamefont {Shabbir}(2010)}]{Shabbir2010}%
  \BibitemOpen
  \bibfield  {author} {\bibinfo {author} {\bibfnamefont {G.}\ \bibnamefont
  {Shabbir}},
  \bibinfo {author} {\bibfnamefont {S.}\ \bibnamefont
  {Khan}},
       \href {\doibase 10.1142/S0217732310033189} {\bibfield  {journal}
  {\bibinfo  {journal} {Mod. Phys. Lett. A}\ }\textbf {\bibinfo {volume} {25}},\ \bibinfo{issueNo.}{25}, \bibinfo{pages}{2145} (\bibinfo {year} {2010})}}\BibitemShut {NoStop}
       \bibitem [{\citenamefont {Ali}(2016)}]{Ali2016}%
  \BibitemOpen
  \bibfield  {author} {\bibinfo {author} {\bibfnamefont {A. T.}\ \bibnamefont
  {Ali}},
  \bibinfo {author} {\bibfnamefont {S.}\ \bibnamefont
  {Khan}} and
  \bibinfo {author} {\bibfnamefont {A.}\ \bibnamefont
  {Alghanemi}},  
        {}{\bibfield  {journal}
  {\bibinfo  {journal} {arXiv:1512.04427v2 [physics.gen-ph]}\ }\textbf\  (\bibinfo {year} {2016})}}\BibitemShut {NoStop}
       \bibitem [{\citenamefont {Shabbir}(2015)}]{Shabbir2015}%
  \BibitemOpen
  \bibfield  {author} {\bibinfo {author} {\bibfnamefont {G.}\ \bibnamefont
  {Shabbir}},
  \bibinfo {author} {\bibfnamefont {M. A.}\ \bibnamefont
  {Shahani}},
       \href {\doibase 10.1007/s10773-015-2532-3} {\bibfield  {journal}
  {\bibinfo  {journal} {Int. J. Theor. Phys.}\ }\textbf {\bibinfo {volume} {54}},\ \bibinfo{issueNo.}{9}, \bibinfo{pages}{3027} (\bibinfo {year} {2015})}}\BibitemShut {NoStop}
        \bibitem [{\citenamefont {Shabbir}(2017)}]{Shabbir2017}%
  \BibitemOpen
  \bibfield  {author} {\bibinfo {author} {\bibfnamefont {G.}\ \bibnamefont
  {Shabbir}},
  \bibinfo {author} {\bibfnamefont {M. A.}\ \bibnamefont
  {Shahani}},
  \bibinfo {author} {\bibfnamefont {M. A.}\ \bibnamefont
  {Qureshi}} and
  \bibinfo {author} {\bibfnamefont {F. M.}\ \bibnamefont
  {Mahomed}},
       \href {\doibase 10.1088/0253-6102/68/5/611} {\bibfield  {journal}
  {\bibinfo  {journal} {Commun. Theor. Phys.}\ }\textbf {\bibinfo {volume} {68}},\ \bibinfo{issueNo.}{5}, \bibinfo{pages}{611} (\bibinfo {year} {2017})}}\BibitemShut {NoStop}
   \bibitem [{\citenamefont {Yano}(1955)}]{Yano1955}%
  \BibitemOpen
  \bibfield  {author} {\bibinfo {author} {\bibfnamefont {K.}\ \bibnamefont
  {Yano}},
       {{}{\bibfield  {chapter}
  {\bibinfo  {chapter} \emph{The theory of Lie derivatives and its Applications
}}}}{\  (\bibinfo{publisher}{North-Holland Pub. Co.},\
  \bibinfo {year} {1955})}
 } \BibitemShut {NoStop}%
          \bibitem [{\citenamefont {McIntosh}(1976)}]{McIntosh1976}%
  \BibitemOpen
  \bibfield  {author} {\bibinfo {author} {\bibfnamefont {C. B. G.}\ \bibnamefont
  {McIntosh}},
    \href {\doibase 10.1007/BF00763435} {\bibfield  {journal}
  {\bibinfo  {journal} {Gen. Relat. Gravit.}\ }\textbf {\bibinfo {volume} {7}},\ \bibinfo{issueNo.}{2}, \bibinfo{pages}{199} (\bibinfo {year} {1976})}}\BibitemShut {NoStop}
          \bibitem [{\citenamefont {Berger}(1976)}]{Berger1976}%
  \BibitemOpen
  \bibfield  {author} {\bibinfo {author} {\bibfnamefont {B. K.}\ \bibnamefont
  {Berger}},
    \href {\doibase 10.1063/1.523052} {\bibfield  {journal}
  {\bibinfo  {journal} {J. Math. Phys.}\ }\textbf {\bibinfo {volume} {17}},\ \bibinfo{pages}{1268} (\bibinfo {year} {1976})}}\BibitemShut {NoStop}
    \bibitem [{\citenamefont {Jaen}(2014)}]{Jaen2014}%
  \BibitemOpen
  \bibfield  {author} {\bibinfo {author} {\bibfnamefont {X.}\ \bibnamefont
  {Ja\'{e}n}},
  \bibinfo {author} {\bibfnamefont {A.}\ \bibnamefont
  {Molina}},
    \href {\doibase 10.1007/s10714-014-1745-8} {\bibfield  {journal}
  {\bibinfo  {journal} {Gen. Relat. Gravit.}\ }\textbf {\bibinfo {volume} {46}},\ \bibinfo{pages}{1745} (\bibinfo {year} {2014})}}\BibitemShut {NoStop}
   \bibitem [{\citenamefont {Gad}(2015)}]{Gad2015}%
  \BibitemOpen
  \bibfield  {author} {\bibinfo {author} {\bibfnamefont {R. M.}\ \bibnamefont
  {Gad}},
    \href {\doibase 10.1007/s10773-015-2528-z} {\bibfield  {journal}
  {\bibinfo  {journal} {Int. J. Theor. Phys.}\ }\textbf {\bibinfo {volume} {54}},\ \bibinfo{issueNo.}{8}, \bibinfo{pages}{2932} (\bibinfo {year} {1976})}}\BibitemShut {NoStop}
    \bibitem [{\citenamefont {Kar}(1996)}]{Kar1996}%
  \BibitemOpen
  \bibfield  {author} {\bibinfo {author} {\bibfnamefont {S.}\ \bibnamefont
  {Kar}},
    \href {\doibase 10.1103/PhysRevD.53.2071} {\bibfield  {journal}
  {\bibinfo  {journal} {Phys. Rev. D}\ }\textbf {\bibinfo {volume} {53}},\ \bibinfo{pages}{2071} (\bibinfo {year} {1996})}}\BibitemShut {NoStop}
     \bibitem [{\citenamefont {Abreu}(2011)}]{Abreu2011}%
  \BibitemOpen
  \bibfield  {author} {\bibinfo {author} {\bibfnamefont {G.}\ \bibnamefont
  {Abreu}},
  \bibinfo {author} {\bibfnamefont {M.}\ \bibnamefont
  {Visser}},
    \href {\doibase 10.1103/PhysRevD.83.104016} {\bibfield  {journal}
  {\bibinfo  {journal} {Phys. Rev. D}\ }\textbf {\bibinfo {volume} {83}},\ \bibinfo{pages}{104016} (\bibinfo {year} {2011})}}\BibitemShut {NoStop}
      \bibitem [{\citenamefont {Harko}(2012)}]{Harko2012}%
  \BibitemOpen
  \bibfield  {author} {\bibinfo {author} {\bibfnamefont {T.}\ \bibnamefont
  {Harko}},
  \bibinfo {author} {\bibfnamefont {F. S. N.}\ \bibnamefont
  {Lobo}},
    \href {\doibase 10.1103/PhysRevD.86.124034} {\bibfield  {journal}
  {\bibinfo  {journal} {Phys. Rev. D}\ }\textbf {\bibinfo {volume} {86}},\ \bibinfo{pages}{124034} (\bibinfo {year} {2012})}}\BibitemShut {NoStop}
      \bibitem [{\citenamefont {Tsagas}(2013)}]{Tsagas2013}%
  \BibitemOpen
  \bibfield  {author} {\bibinfo {author} {\bibfnamefont {C. G.}\ \bibnamefont
  {Tsagas}},
  \bibinfo {author} {\bibfnamefont {M. I.}\ \bibnamefont
  {Kadiltzoglou}},
    \href {\doibase 10.1103/PhysRevD.88.083501} {\bibfield  {journal}
  {\bibinfo  {journal} {Phys. Rev. D}\ }\textbf {\bibinfo {volume} {88}},\ \bibinfo{pages}{083501} (\bibinfo {year} {2013})}}\BibitemShut {NoStop}
       \bibitem [{\citenamefont {Das}(2014)}]{Das2014}%
  \BibitemOpen
  \bibfield  {author} {\bibinfo {author} {\bibfnamefont {S.}\ \bibnamefont
  {Das}},
    \href {\doibase 10.1103/PhysRevD.89.084068} {\bibfield  {journal}
  {\bibinfo  {journal} {Phys. Rev. D}\ }\textbf {\bibinfo {volume} {89}},\ \bibinfo{pages}{084068} (\bibinfo {year} {2014})}}\BibitemShut {NoStop}
  \bibitem [{\citenamefont {Fathi}(2016)}]{Fathi2016}%
  \BibitemOpen
  \bibfield  {author} {\bibinfo {author} {\bibfnamefont {M.}\ \bibnamefont
  {Fathi}},
  \bibinfo {author} {\bibfnamefont {M.}\ \bibnamefont
  {Mohseni}},
    \href {\doibase 10.1140/epjp/i2016-16360-7} {\bibfield  {journal}
  {\bibinfo  {journal} {Eur. Phys. J. Plus}\ }\textbf {\bibinfo {volume} {131}},\ \bibinfo{pages}{360} (\bibinfo {year} {2016})}}\BibitemShut {NoStop}
  \bibitem [{\citenamefont {Thompson}(2017)}]{Thompson2017}%
  \BibitemOpen
  \bibfield  {author} {\bibinfo {author} {\bibfnamefont {R. T.}\ \bibnamefont
  {Thompson}},
  \bibinfo {author} {\bibfnamefont {M.}\ \bibnamefont
  {Fathi}},
       \href {\doibase 10.1103/PhysRevD.96.105006} {\bibfield  {journal}
  {\bibinfo  {journal} {Phys. Rev. D}\ }\textbf {\bibinfo {volume} {96}}, \ \bibinfo {No.} {105006} (\bibinfo {year} {2017})}}\BibitemShut {NoStop}%
  \bibitem [{\citenamefont {Sotiriou}(2010)}]{Sotiriou2010}%
  \BibitemOpen
  \bibfield  {author} {\bibinfo {author} {\bibfnamefont {T. P.}\ \bibnamefont
  {Sotiriou}},
  \bibinfo {author} {\bibfnamefont {V.}\ \bibnamefont
  {Faraoni}},
       \href {\doibase 10.1103/RevModPhys.82.451} {\bibfield  {journal}
  {\bibinfo  {journal} {Rev. Mod. Phys.}\ }\textbf {\bibinfo {volume} {82}}, \ \bibinfo {No.} {451} (\bibinfo {year} {2010})}}\BibitemShut {NoStop}%
  \bibitem [{\citenamefont {Frankel}(2012)}]{Frankel2012}%
  \BibitemOpen
  \bibfield  {author} {\bibinfo {author} {\bibfnamefont {T.}\ \bibnamefont
  {Frankel}},
       {\href {\doibase 10.1017/CBO9781139061377} {\bibfield  {chapter}
  {\bibinfo  {chapter} \emph{The Geometry of Physics:
An Introduction}}}}
  {
  \  (\bibinfo  {publisher} {Cambridge University Press},\
  \bibinfo {year} {2012})}
 } \BibitemShut {NoStop}%
  \bibitem [{\citenamefont {Chandrasekhar}(1998)}]{Chandrasekhar1998}%
  \BibitemOpen
  \bibfield  {author} {\bibinfo {author} {\bibfnamefont {S.}\ \bibnamefont
  {Chandrasekhar}},
       {{}{\bibfield  {chapter}
  {\bibinfo  {chapter} \emph{The Mathematical Theory of Black Holes
}}}}{\  (\bibinfo{publisher}{Oxford University Press},\
  \bibinfo {year} {1998})}
 }\BibitemShut {NoStop}%
  \bibitem [{\citenamefont {Fathi}(2013)}]{Fathi2013}%
  \BibitemOpen
  \bibfield  {author} {\bibinfo {author} {\bibfnamefont {M.}\ \bibnamefont
  {Fathi}},
     \href {\doibase 10.1088/1674-1137/37/2/025101} {\bibfield  {journal}
  {\bibinfo  {journal} {Chin. Phys. C}\ }\textbf {\bibinfo {volume} {37}}, \ \bibinfo {No.}{2}, \bibinfo {pages} {025101} (\bibinfo {year} {2013})}}\BibitemShut {NoStop}%
  \end{thebibliography}

%
\end{document}